\documentclass[twocolumn,aip,reprint,numerical]{revtex4-1}
\usepackage{array}
\usepackage{float}
\usepackage{textcomp}
\usepackage{bm}
\usepackage{tipa}
\usepackage{amsmath}
\usepackage{amssymb}
\usepackage{graphicx}
\usepackage{color}

\makeatletter

\newcommand{\lyxmathsym}[1]{\ifmmode\begingroup\def\b@ld{bold}
	\text{\ifx\math@version\b@ld\bfseries\fi#1}\endgroup\else#1\fi}

\usepackage{dcolumn}
\usepackage{bm}
\usepackage{epsfig}
\usepackage{braket}

\usepackage{lineno}
\makeatother

\begin{document}
\title{Non-Markovian recovery makes complex networks more resilient against large-scale failures}

\author{Zhao-Hua Lin}
\affiliation{State Key Laboratory of Precision Spectroscopy and School of Physics and Electronic Science,
East China Normal University, Shanghai 200241, China}

\author{Mi Feng}
\affiliation{Shanghai Key Laboratory of Multidimensional Information Processing, East China Normal University, Shanghai 200241, China}

\author{Ming Tang}\email{tangminghan007@gmail.com}
\affiliation{State Key Laboratory of Precision Spectroscopy and School of Physics and Electronic Science,
East China Normal University, Shanghai 200241, China}
\affiliation{Shanghai Key Laboratory of Multidimensional Information Processing, East China Normal University, Shanghai 200241, China}

\author{Zonghua Liu}\email{zhliu@phy.ecnu.edu.cn}
\affiliation{State Key Laboratory of Precision Spectroscopy and School of Physics and Electronic Science,
East China Normal University, Shanghai 200241, China}

\author{Chen Xu}
\affiliation{School of Physical Science and Technology, Soochow
University, Suzhou 215006, China}

\author{Pak Ming Hui}
\affiliation{Department of Physics, The Chinese University of Hong
Kong, Shatin, Hong Kong SAR, China}

\author{Ying-Cheng Lai}
\affiliation{School of Electrical, Computer and Energy Engineering, Arizona State University, Tempe, AZ 85287, USA}

\begin{abstract}
\section*{Abstract} \label{sec:abstract}

Non-Markovian spontaneous recovery processes with a time delay (memory) are ubiquitous
in the real world. How does the non-Markovian characteristic affect failure propagation in
complex networks? We consider failures due to internal causes at the nodal level and
external failures due to an adverse environment, and develop a pair approximation analysis
taking into account the two-node correlation. In general, a high failure stationary state can
arise, corresponding to large-scale failures that can significantly compromise the functioning
of the network. We uncover a striking phenomenon: memory associated with nodal recovery
can counter-intuitively make the network more resilient against large-scale failures. In natural
systems, the intrinsic non-Markovian characteristic of nodal recovery may thus be one reason
for their resilience. In engineering design, incorporating certain non-Markovian features into
the network may be beneficial to equipping it with a strong resilient capability to resist
catastrophic failures.

\end{abstract}
\maketitle

\section*{Introduction} \label{sec:intro}

The dynamics of failure propagation on complex networks constitute an
active area of research in network science and engineering with significant
and broad applications. This is because the functioning of a modern society
relies on the cooperative working of many networked systems such as the
electrical power grids, various transportation networks, computer and
communication networks, and business networks, but these networks typically
possess a complex structure and are vulnerable to failures and intentional
attacks. Among the diverse failure scenarios, one of the most severe types
is cascading failures~\cite{ML:2002}, where the failure of some nodes would
cause their neighbors to fail and the process would propagate to the entire
network, disabling a large fraction of the nodes and causing
malfunctioning of the system at a large scale~\cite{ZPL:2004,ZPLY:2005,
GC:2007,Bialek:2007,dobson2007complex,Gleeson:2008,rosato2008modelling,
HLC:2008,SBPBH:2008,YWLC:2009,takayasu2010econophysics,HL:2011,WLA:2011,
LWLW:2012}. Classic examples of cascading failures include power blackout
- the collapse of power grids~\cite{Bialek:2007,dobson2007complex},
traffic jams~\cite{li2015percolation}, and even economic
depression~\cite{parshani2011critical,WLA:2011}. Previous studies mostly
focused on how cascading failures occur, how network structures and failure
propagation are related, and on network robustness and vulnerability to
failure propagation~\cite{watts2002simple,dodds2004universal,
simonsen2008transient,buldyrev2010catastrophic,ganin2017resilience}.

A tacit assumption employed in many previous studies of cascading failures is
irreversible failure propagation, where a node, if it has failed, cannot
recover and is no longer able to function actively. A failed node is then
removed from the network completely, including all the links associated with
it. There are real-world situations of networked systems, such as financial
and transportation networks, where failed nodes can recover from
malfunctioning spontaneously after a collapse~\cite{nudo2013recovery,
shang2015impact,hu2016recovery,white2001autonomic,toohey2007self,
desmurget2007contrasting}. In general, there are
two types of failure-and-recovery scenarios~\cite{majdandzic2014spontaneous}:
internal and external. In the first type, a node fails because of internal
causes (e.g., the occurrence of some abnormal or undesired dynamical behaviors
within the node), which is independent of the states of its neighbors. In this
case, the node can recover spontaneously after a period of time. An example
is the failure of a company characterized by a drop in its market value due
to poor management, followed by recovery due to internal restructuring. The
second type is external failures, where a node's failure is externally
triggered, e.g., by the failures of its neighboring nodes. After a period of
time, as its local ``environment'' is improved, the node is able to recover
spontaneously. The time of recovery depends not only on the specific type
of failure-and-recovery mechanism, i.e., whether internal or external, but
also on the individual node and its position in the network. For example,
for a given node in the network, it may take longer to complete an internal
restructuring process to recover from a failure due to an internal than an
external cause. Previous computation and mean-field analysis have revealed
that cascading dynamics incorporating a failure-and-recovery mechanism can
exhibit a rich variety of phenomena such as phase transitions, hysteresis,
and phase flipping~\cite{majdandzic2014spontaneous,podobnik2014network,
Podobnik2015,Podobnik2015a,Majdandzic2016}.
With respect to the resilience responses of networks, the effects of removing
a fraction of nodes and links on network functions were
studied~\cite{national2012disaster,gao2016universal,ganin2016operational,
linkov2019science}, demonstrating that resilience can be used to characterize
the critical functionality of the network with applications in complex
infrastructure engineering~\cite{ganin2016operational,linkov2019science}.

In spite of the variations in the recovery dynamics across networks or even
nodes in the same network, generally the process can be classified into
two distinct types: Markovian and non-Markovian. In a Markovian recovery
process, an event occurs at a fixed rate and the inter-event time follows
an exponential distribution~\cite{pastor2015epidemic,wang2017unification,
de2018fundamentals}, rendering memoryless the process. On the contrary, a
non-Markovian recovery process has memory, as the current state of a node
depends not only on the most recent state but also on the previous states.
In this case, the inter-event time distribution is not exponential but
typically exhibits a heavy tail. For example, in human activity and
interaction dynamics, the occurrences of contacts among the individuals in
a social network can be characteristically non-Markovian, for which there
is mounting empirical evidence~\cite{Barabasi:2005,GHB:2008,
SGMB:2012,ZYZZHL:2013,ZHHLL:2014,PSRPGB:2015,ZZYWL:2016,YWGL:2017}.
Non-Markovian type of recovery process also occurs in biochemical
reactions~\cite{bratsun2005delay} and in the financial
markets~\cite{takayasu2010econophysics,scalas2006waiting}. We note
that, in the context of spreading dynamics on complex networks, the effects of
the non-Markovian process, due to its high relevance to the real world,
have attracted growing attention~\cite{VRLB:2007,IM:2009,MB:2013,JPKK:2014,
KRV:2015,SGB:2017,SMBK:2018,FCTL:2019}. From
the point of view of mathematical analysis, incorporating memories into
the dynamical process makes analytic treatment challenging.

While the impacts of non-Markovian processes on spreading dynamics
have been reasonably well documented~\cite{VRLB:2007,IM:2009,MB:2013,JPKK:2014,
KRV:2015,SGB:2017,SMBK:2018,FCTL:2019},
there has been little work so far addressing the influence of non-Markovian
recovery process on failure propagation dynamics. In this paper, we address
this issue systematically through a comparison study of two types of dynamical
processes: one with Markovian and another with non-Markovian recovery. In
the Markovian recovery (MR) model, failures due to internal and external
causes will recover with different constant rates. In the non-Markovian
recovery (NMR) model, such a constant rate cannot be defined. We thus
resort to the recovery time. In particular, we assume that the failed nodes
due to internal and external causes will take different time to recover, so a
memory effect is naturally built into the model. For each model, we
develop a mean-field theory and an analysis based on the pair
approximation (PA)~\cite{majdandzic2014spontaneous,Valdez2016,Bottcher2017,
Bottcher2017a,keeling1997correlation} that retains the two-node correlation
but ignores any correlation of higher orders. Comparing results with numerical
simulations indicates that both mean-field theory and PA analysis capture
the key features of the failure propagation dynamics qualitatively, but the
PA analysis yields results that are in better quantitative agreement with
numerics. The counterintuitive and striking phenomenon is then that 
non-Markovian character with a memory effect makes the network more resilient 
against large-scale failures. There are two implications. Firstly,
in physical, biological, or other natural networked systems, the
intrinsic non-Markovian character of nodal recovery may be one reason for
resilience of these networks and their existence in a harsh environment.
Secondly, in engineering and infrastructure design, incorporating certain
non-Markovian features into the network may help strengthen its resilience
and robustness.

\section*{Results}

\subsection{Spontaneous recovery models} \label{sec:model}

For general failure propagation dynamics on a network, a node can be in one
of two states: an active (labeled as $A$-type) state in which the node
functions properly and an inactive state ($I$-type) in which the node
has failed. To distinguish the causes for a node to become inactive, we
label an inactive node due to internal or external failure as $X$-type
or $Y$-type, respectively.

\begin{figure}
\includegraphics[width=1.0\linewidth]{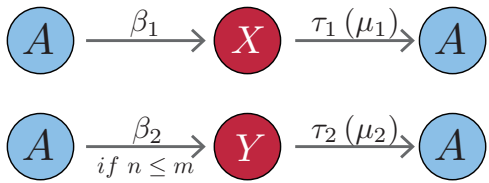}
\caption{ {\bf Schematic illustration of NMR and MR models}.  
An active ($A$-type) node may fail spontaneously at the
rate $\beta_1$ to become an $X$-type node due to internal causes, or it may
fail at the rate $\beta_2$ to become a $Y$-type node when the number of its
$A$-type neighbors $n$ is less than or equal to a threshold $m$ setting the
necessary neighboring support for the proper functioning of a node. In the
NMR model, the $X$-type and $Y$-type nodes take the time duration $\tau_1$
and $\tau_2$ from the time they are generated to recover, respectively. In
the MR model, the $X$-type and $Y$-type nodes recover respectively at the
rates $\mu_1$ and $\mu_2$.}
\label{fig:model}
\end{figure}

In the NMR model, an $A$-type node may fail spontaneously at the rate
$\beta_1$ to become an $X$-type node; or it may fail at the rate $\beta_2$ to
become a $Y$-type node when the number of its $A$-type neighboring nodes is
less than or equal to a threshold integer value $m$ that sets the limit on
neighboring support for proper functioning of a node. Without loss of
generality, we assume that external failures occur more frequently than
internal failures: $\beta_1 < \beta_2$. This is often the case as
internal failures can be made less probable by building up the capability
of the nodes through better equipment and/or management, while external
failures are uncontrollable and more difficult to avoid.
For examples, falling stocks may be the result of unanticipated changes
in the market rather than poor management. In a road network, failures are
caused more often by congestion than by physical failures.
Once a node becomes inactive, it takes time $\tau_{1}$ to recover from an
internal failure (when the node is of the $X$-type) or time $\tau_{2}$ to
recover from an external failure (when the node is of the $Y$-type). The
non-Markovian characteristic is taken into account through the incorporation
of a memory effect into the model. In particular, the nodes that will recover
at time $t$ constitute those that were turned into $X$-type inactive nodes at
the time $t-\tau_1$ and those turned into $Y$-type inactive nodes at the time
$t-\tau_2$. Here, we assume $\tau_{1} > \tau_{2}$, for the reason that
repairing a node or restructuring the management due to the malfunctioning
of the node itself would need more time.
For example, reorganizing a company or repairing a road often takes more time.
The failure processes characterized by the rates $\beta_1$ and $\beta_2$ as
well as the recovery processes as determined by $\tau_1$ and $\tau_2$ are
schematically illustrated in Fig.~\ref{fig:model}.

Note that the case of $\tau_1<\tau_2$ may also arise in the real world. For
example, for an infrastructure network in civil engineering, when an
earthquake strikes and destroys buildings (nodes), the time to rebuild can
be longer than that required for recovering from internal failures, e.g.,
the collapse of a roof due to some material failure. Our computations of
this case yield qualitatively similar results to those in the case of
$\tau_1>\tau_2$ - see Supplementary Note 3 for detail.

The MR and NMR models differ only in the recovery processes. In the MR model,
an inactive node of the $X$-type or the $Y$-type recovers at a constant rate
$\mu_1$ or $\mu_2$, respectively, as illustrated in Fig.~\ref{fig:model}.
Consequently, the number of nodes recovered at time $t$ depends only on the
number of inactive nodes of both $X$-type and $Y$-type at the previous
time step.

To develop theories for failure propagation on networks with MR or NMR
recovery process and to identify the key differences between the two type
of dynamics, we focus on random regular networks. In the numerical
simulations, we use a relatively large network size $N=3\times 10^4$ with
the degree $k=35$. In the NMR model, the recovery times are taken to be
$\tau_1=100$ and $\tau_2=1$ for the $X$-type and $Y$-type of nodes,
respectively. In the MR model, the values of the recovery rates are set to be
$\mu_1=1/\tau_1=0.01$ and $\mu_2=1/\tau_2=1$ so that they correspond to the
same scales for the recovery times in the NMR model (see
Supplementary Note 1 for a more detailed explanation). The threshold
values in both models are $m=15$. Synchronous updating is invoked in
simulations with the time step $\Delta{t}=0.01$.

\subsection{Markovian Recovery Process} \label{sec:MRP}

\begin{figure*}
\includegraphics[width=1.0\linewidth]{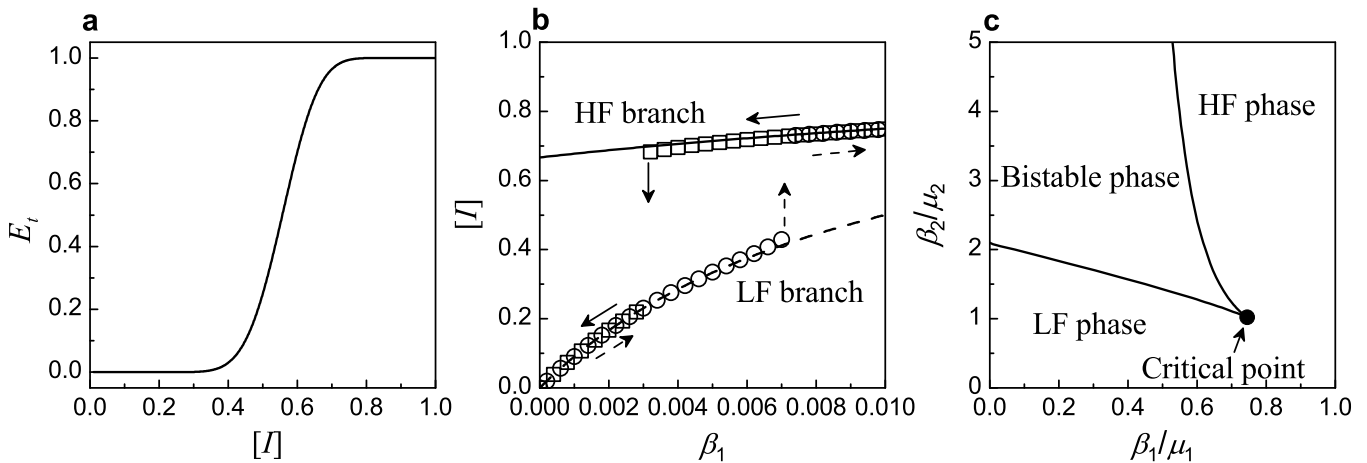}
\caption{ {\bf Behaviors of MR model}. (a) Probability $E_t([I])$ in
the mean-field theory [Eq.~(\ref{eq:mfr_mf_e})] as a function of $[I]$.
(b) Dependence of $[I]$ on $\beta_1$ in the steady state for $\beta_2=2.0$,
$\mu_1=0.01$, and $\mu_2=1$. The high-failure (solid curve) and low-failure
(dashed curve) branches are calculated by the mean-field theory. The
simulation data are obtained by swabbing $\beta_1$ up and down in step
of $0.002$, starting with $[I]_{0} = 0$ for $\beta_1=0$. The final state of
a value of $\beta_1$ is used as the initial state of the simulations for
the next value of $\beta_1$. The arrows indicate the simulation results
when the value of $\beta_1$ moves up and down. (c) Phase diagram on the
$\beta_2/\mu_2$-$\beta_1/\mu_1$ plane as predicted by the mean-field theory.
Systems in the bistable phase will evolve either to a high-failure or a
low-failure phase depending on the initial conditions. Beyond the critical
point, there is no distinction between the low-failure and high-failure
phases.}
\label{fg:X2}
\end{figure*}

{\bf Mean-field theory}. 
We start with setting up the dynamical equations for MR dynamics and comparing
results with simulations. Based on the mean-field theory in part {\bf A} of 
{\it Methods}, we first examine the behavior of $E_{t}([I])$ in 
Eq.~(\ref{eq:mfr_mf_e}). Figure~\ref{fg:X2}(a) shows the dependence of 
$E_{t}([I])$ on the fraction of failed nodes $[I]$. It can be seen that 
$E_{t}$ exhibits two different types of behaviors over a large part of $[I]$: 
$E_{t} \sim 0$ for a wide range of small $[I]$ values (low failure) and 
$E_{t} \sim 1$ for a range of large $[I]$ values (high failure). In the low 
failure state, external failure events rarely occur. In the high failure 
state, an active node is supported by an insufficient number of active 
neighbors and external failure events almost always happen. It implies that 
the stationary state $[I]$ can possess two branches: setting $E_{t}=0$ in 
Eq.~(\ref{eq:selfI}) gives $[I]=1-1/(\beta_1/\mu_1+1)$ as the low-failure 
branch, while setting $E_{t}=1$ gives $[I]=1-1/(\beta_2/\mu_2+\beta_1/\mu_1+1)$
as the high-failure branch. The two branches are shown in Fig.~\ref{fg:X2}(b) 
(dashed and solid curves) in terms of the dependence of $[I]$ on $\beta_{1}$, 
for $\mu_1=0.01$, $\beta_2=2$ and $\mu_2=1$ as an example. To check which 
branch the system would take on and whether there are two states for some 
range of parameters, the simulation results for moving the value of 
$\beta_{1}$ up (circles) and down (squares) are shown in Fig.~\ref{fg:X2}(b) 
for comparison. As the values of $\beta_{1}$ are increased or decreased, the 
initial state is taken to be the final state corresponding to the previous 
value of $\beta_{1}$ - the adiabatic process. The results indicate that: 
(i) the values of $[I]$ from simulations follow the two branches given by 
the mean-field approximation, and (ii) the low-failure (high-failure) branch 
is followed when moving $\beta_{1}$ up (down) until a particular value at 
which there is a jump to the high-failure (low-failure) branch - the signature 
of a hysteresis. The results also imply that if one starts from the initial 
conditions $[X]_{0} \neq 0$ and $[Y]_{0}=0$, there exists a critical value 
of $\beta_\mathrm{c} \approx 0.007$ for a sudden increase in the number of 
failed nodes when $[X]_{0}$ is small as the system will follow the low-failure 
branch. However, for large $[X]_{0}$, the critical value $\beta_\mathrm{c}$ 
becomes $\beta_\mathrm{c} \approx 0.003$ as the system will follow
the high-failure branch. A plot of $\beta_\mathrm{c}$ against $[X]_{0}$ will
therefore exhibit two plateaus with $\beta_\mathrm{c} \approx 0.007$ for small
$[X]_{0}$ and $\beta_\mathrm{c} \approx 0.003$ for large $[X]_{0}$.

The mean-field approximation not only simplifies the analysis but also
provides insights into the dynamical process. For example, the mean-field
theory suggests the ratios $\beta_{1}/\mu_{1}$ and $\beta_{2}/\mu_{2}$ as
key parameters. In general, solutions can be obtained numerically by solving
Eq.~(\ref{eq:selfI}) together with Eq.~(\ref{eq:mfr_mf_e}). The results are
shown in Fig.~\ref{fg:X2}(c) as a phase diagram. For parameters falling into
the regions corresponding to the low-failure (high-failure) phase, the system
will evolve into a low-failure (high-failure) state. For parameters in the
bistable phase, the system will evolve either to a low-failure or a
high-failure state, depending on the initial conditions. The high-failure and
low-failure phase boundaries meet at the critical point determined by
$\beta_1/\mu_1\approx 0.745$ and $\beta_2/\mu_2\approx 1.020$.

\begin{figure*}
\includegraphics[width=0.8\linewidth]{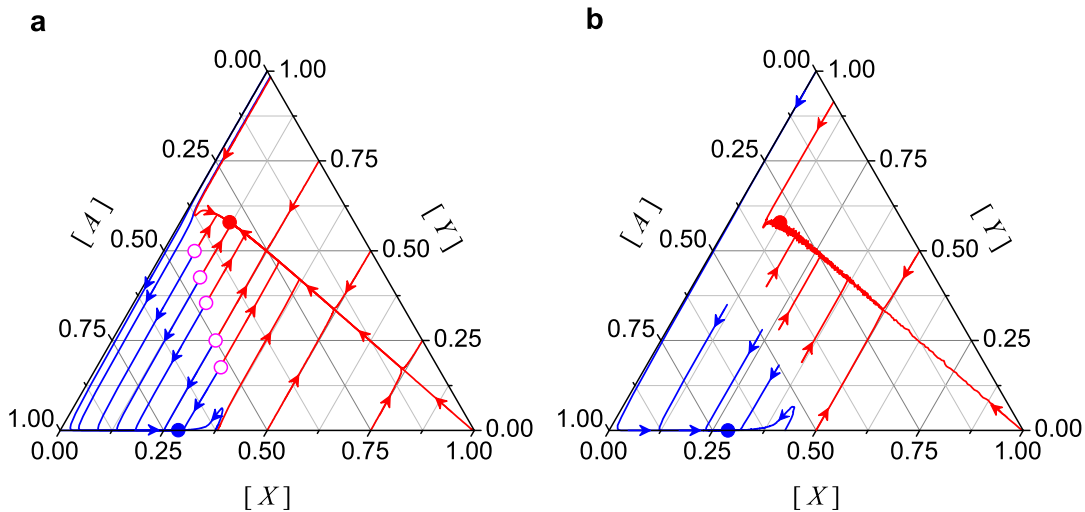}
\caption{ {\bf Evolutionary properties of MR dynamics}. The dynamical
process can be represented as a flow diagram with the lines showing how the
fractions of $[X]$, $[Y]$, and $[A]$ evolve in time. The solid circles are
the fixed points that the system will finally evolve into. (a) Theoretical
calculations based on the mean-field theory. The open circles trace out a
separatrix, with systems on different sides evolving into different fixed
points. (b) Simulation results. The system parameters are $\beta_1=0.004$,
$\beta_2=2.0$, $\mu_1=0.01$, and $\mu_2=1$.}
\label{fg:X3}
\end{figure*}

In addition to the stationary state, the evolution of the system can also be
studied by iterating Eqs.~(\ref{eq:mfr_x}) and (\ref{eq:mfr_y}) for a given
initial condition. Figure~\ref{fg:X3}(a) shows the evolution of the MR dynamics
as obtained by the mean-field theory for $\beta_1=0.004$ and $\beta_2=2$. In
the three-dimensional space formed by $[A]$, $[X]$, and $[Y]$, the sum rule
$[A]+[X]+[Y]=1$ defines a triangular plane, as shown in Fig.~\ref{fg:X3}. At
any time $t$, the state of the system is characterized by a point in the
plane. The results show that the MR dynamics will evolve into
either the low-failure or the high-failure state (filled circles),
depending on where the system begins. The mean-field theory also gives a
separatrix, the line traced out by the open circles, where the system will
evolve into a different state starting from a point on a different side of the
separatrix. For $[X]_0>0.38$, the system will evolve to a high-failure state
with ($[X]$,$[Y]$,$[A]$) given approximately by ($0.119$, $0.580$, $0.301$).
For $[X]_0<0.38$, the system may evolve to the high-failure state or a
low-failure state at around ($0.285$, $0$, $0.715$). Numerical results
are shown in Fig.~\ref{fg:X3}(b), verifying all the features predicted by the
mean-field theory. For example, the high-failure state is given by
($[X]$,$[Y]$,$[A]$) $\sim$ ($0.124$, $0.579$, $0.298$) and the low-failure
state at around ($0.287$, $0$, $0.713$), both are quite close to the values
predicted by the mean-field theory.

{\bf Pairwise approximation theory for the MR model}.
It is possible to formulate a theory that takes into account of two-node
spatial correlation based on the pairwise approximation (PA). The basic idea 
is to follow the evolution of different types of links, i.e., links that 
connect different pairs of neighboring nodes~\cite{keeling1997correlation}.
The PA method has been used widely in studying epidemic and information 
spreading~\cite{ben1992mean,mata2013pair,gross2006epidemic}, and in coevolving 
voter models and adaptive games with two or more 
strategies~\cite{ji2013coevolve,zhang2013SG,zhang2014coevolve,choi2017coevolve}.
In Part {\bf B} of {\it Methods}, we develop a PA based theory for the MR model.

\begin{figure*}
\includegraphics[width=1.0\linewidth]{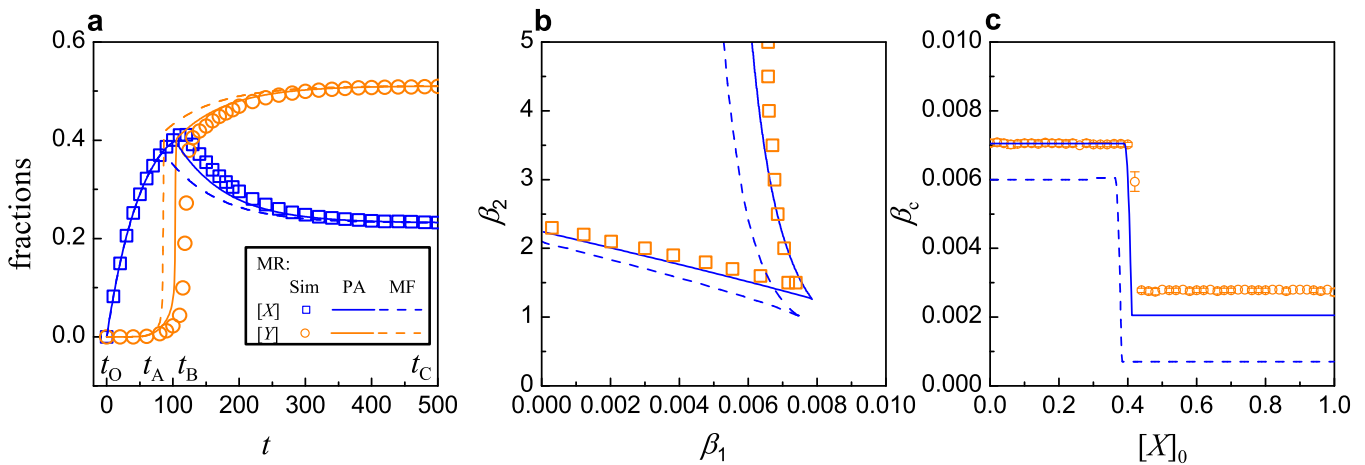}
\caption{ {\bf Comparison of simulation results with predictions from PA
analysis and mean-field theory for the MR model}. (a) Time evolution of the
fraction of inactive nodes. The initial conditions are $[X]_0=[Y]_0=0$. The
parameter values are $\beta_1=0.009$, $\beta_2=2.0$, $\mu_1=0.01$, and
$\mu_2=1.0$. Several time instants are marked to facilitate visualization
of the time evolution in different stages: $t_\mathrm{O}=0$, 
$t_\mathrm{A}=60$, $t_\mathrm{B}=120.7$, and $t_\mathrm{C}=480$. 
(b) Phase diagram on the ($\beta_2$-$\beta_1$) parameter
plane. Other parameters are $\mu_1=0.01$, and $\mu_2=1.0$. (c) Dependence
of $\beta_\mathrm{c}$ on the initial value of $[X]_{0}$, with $[Y]_{0}=0$. The
solid (dashed) curve is calculated by the PA (mean-field) theory, and the
symbols are for simulation results obtained by averaging over $100$
realizations.}
\label{fg:X5}
\end{figure*}

Figure~\ref{fg:X5} presents a comparison of the predictions of the PA analysis
and mean-field theory with the numerical results, where Fig.~\ref{fg:X5}(a)
shows the time evolution of $[X]_{t}$ and $[Y]_{t}$ from the initial state
$[X]_0=[Y]_0=0$ for $\beta_1=0.009$, $\beta_2=2.0$, $\mu_1=0.01$, and $\mu_2=1$.
While both mean-field and PA theories capture the key features in time
evolution, the results of PA are in better agreement with those from
simulations. It is useful to understand the dynamical behaviors in the MR
model qualitatively (so as to enable a meaningful comparison with those of the
NMR model later). For this purpose, we identify several stages in the time
evolution as marked in Fig.~\ref{fg:X5}(a). In the early stage, i.e.,
$t \in [t_\mathrm{O},t_\mathrm{A}]$, most nodes are active and they have more 
active neighbors, violating the condition $n_{A} \leq{m}$. As a result, only 
internal failures occur and $[X]_{t}$ grows but $[Y]_{t}$ decreases and 
eventually vanishes. For $t \in [t_\mathrm{A},t_\mathrm{B}]$, $[X]_{t}$, 
active nodes start to fail into $Y$-type nodes, leading to fewer active 
nodes in the system and triggering more external nodal failures. This results 
in the observed rapid increase in $[Y]$. In the later stage 
$t \in [t_\mathrm{B},t_\mathrm{C}]$, there are more failed
nodes than active ones. While the failed nodes of $X$ and $Y$ types can
recover with their respective rates, the remaining or recovered active nodes
will more likely fail again through external than internal causes
due to the many failed nodes surrounding the active nodes. Consequently,
in this later stage, $[Y]_{t}$ increases and $[X]_{t}$ decreases
toward their respective steady-state values for $t\rightarrow\infty$, with
$[Y] > [X]$ when the system evolves into a high-failure state. The PA analysis
captures the behavior of $[X]_{t}$ over time and the onset of $[Y]_{t}$ better
than the mean-field analysis. Figure~\ref{fg:X5}(b) shows the phase diagram
for $\mu_1=0.01$ and $\mu_2=1.0$. The mean-field phase diagram is the same as
that shown in Fig.~\ref{fg:X2}(c), where it can be seen that the results of
the PA analysis (solid curves) are indeed in better agreement with the
simulation results than the predictions of the single-node mean-field theory.

Note that Fig.~\ref{fg:X2} reveals the emergence of a critical value of
$\beta_\mathrm{c}$ in the spontaneous failure rate beyond which the system 
incurs a large-scale failure starting from the initial conditions 
$[X]_{0} \neq 0$ and $[Y]_{0}=0$. The critical rate $\beta_\mathrm{c}$ is 
calculated by starting the system from the initial conditions for different 
values of $\beta_{1}$ (for a fixed value of $\beta_2 = 2.0$) and search for 
the value of $\beta_1$ beyond which the system attains a high-failure state 
(see Supplementary Figure 1 in Supplementary Note 2). The critical value thus 
depends on $[X]_{0}$, the initial fraction of failed nodes due to an internal 
mechanism. Figure~\ref{fg:X5}(c) shows the numerically obtained functional 
relation $\beta_\mathrm{c}([X]_{0})$ (open circles), together with two types 
of theoretical prediction (PA analysis and mean-field theory). As the initial 
fraction $[X]_{0}$ is increased from a near zero value, $\beta_\mathrm{c}$ 
maintains at a relatively higher constant value (about 0.007). As $[X]_{0}$ 
increases through the value of about 0.4, the value of the critical rate 
suddenly decreases to about 0.003. We see that, again, the prediction of the 
abrupt change in $\beta_\mathrm{c}$ by the PA analysis is more accurate than 
that by the mean-field theory.

What is the physical meaning of the abrupt decrease in the critical value
of the spontaneous failure rate as displayed in Fig.~\ref{fg:X5}(c)? A higher
value of $\beta_\mathrm{c}$ means that the network system is more resilient to
large-scale failures as it requires a larger rate value to drive the system
into a high-failure state. As the fraction of initially failed nodes is
increased, the network as a whole is more prone to large-scale failure so
we expect the value of $\beta_\mathrm{c}$ to decrease. Because of the lack of
any memory effect in the ideal, Markovian type of recovery process, i.e.,
after a node fails, it either recovers instantaneously or does not recover
(with probabilities determined by the rate of recovery), we expect a
characteristic change in the system dynamics as characterized by the value of
the critical rate $\beta_\mathrm{c}$ to occur in an abrupt manner. Indeed,
as Fig.~\ref{fg:X5}(c) reveals, as the fraction of initially failed nodes
is increased through a threshold value, there is a sudden decrease of about
$50\%$ in the value of $\beta_\mathrm{c}$, giving rise to a first-order type of
transition. This behavior of abrupt transition may not occur in reality
because of the assumed Markovian recovery process, which is ideal and
cannot be expected to arise typically in the physical world. In the next
section, we will demonstrate that making the dynamics more physical by
assuming non-Markovian type of recovery process will drastically alter
the picture of transition in Fig.~\ref{fg:X5}(c).\\

\subsection{Non-Markovian recovery process}

\begin{figure*}
\includegraphics[width=1.0\linewidth]{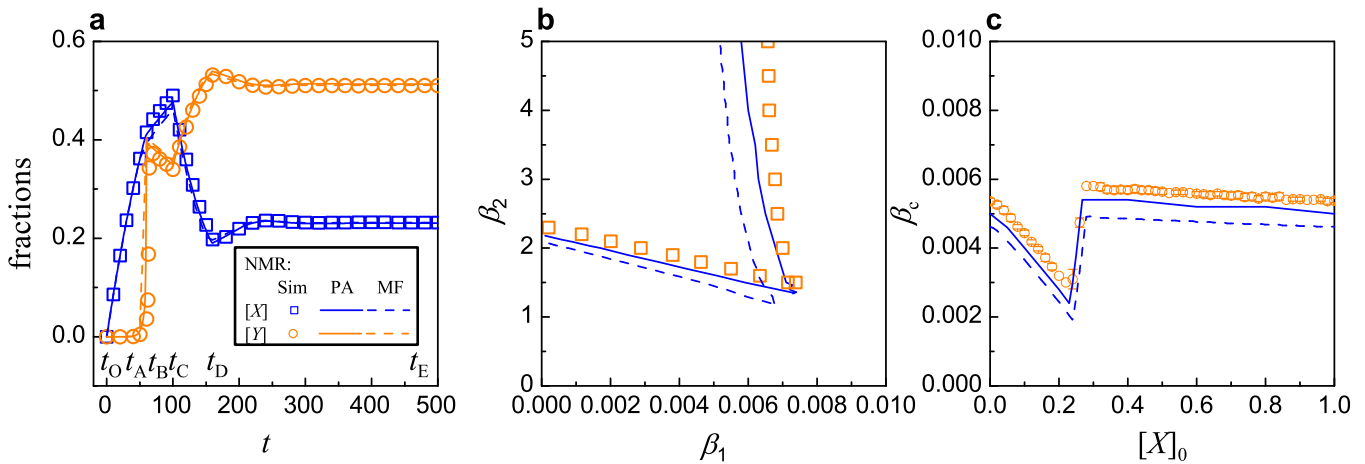}
\caption{ {\bf Benefit of non-Markovian recovery to making the
network more resilient against large-scale failures}. Shown are results from
the NMR model through simulations, PA analysis, and mean-field theory.
(a) Time evolution of inactive nodes from the initial conditions
$[X]_0=[Y]_0=0$, for $\beta_1=0.009$, $\beta_2=2.0$, $\tau_{1} = 100$ (thus
$\mu_1=0.01$), and $\tau_{2} = 1.0$ (thus $\mu_2=1.0$). A number of time
instants are marked for better visualization of the time evolution in
different stages: $t_\mathrm{O}=0$, $t_\mathrm{A}=43.51$, $t_\mathrm{B}=64.68$, $t_\mathrm{C}=100$, $t_\mathrm{D}=164.51$, and $t_\mathrm{E}=480$. 
(b) Phase diagram in the ($\beta_2$-$\beta_1$) parameter plane
for $\tau_1=100$ and $\tau_2=1.0$. The symbols are numerical results, and the
solid and dashed curves are obtained from the PA analysis and mean-field
theory, respectively. (c) Dependence of $\beta_\mathrm{c}$ for reaching a
high-failure state on the initial value of $[X]_{0}$ with $[Y]_{0} = 0$.
The error bars with the simulation results are about $6 \times 10^{-5}$,
which are obtained by averaging over $100$ realizations.}
\label{fg:X6}
\end{figure*}

To analyze failure propagation dynamics in systems with NMR, a viable approach
is to construct difference equations that relate the fractions of types of
nodes and links at time $t+\Delta t$ to those at time $t$. It is necessary to
keep track of the time when a node becomes the $X$ or $Y$ type as well as the
time at which a link becomes type $UV$. In Part {\bf C} of {\it Methods}, 
we develop a PA analysis for the NMR model.
Figure~\ref{fg:X6} shows the simulation results from the NMR model, together
with predictions of the PA analysis and mean-field approximation for
$\Delta t = 0.01$. The time evolution of $[X]_{t}$ and $[Y]_{t}$ is shown
for the parameter setting $\beta_1=0.009$, $\beta_2=2.0$, $\tau_1=100$ (thus
$\mu_{1}=0.01$), and $\tau_2=1$ (thus $\mu_{2}=1$). The initial conditions
are $[X]_0=[Y]_0=0$. Both theories capture the key features of the dynamics.
Comparing with results from the MR model [e.g., Fig.~\ref{fg:X5}(a)], we
see that the time evolution of the dynamical variables in the NMR model is
different from that in the MR model, in spite of the approximately identical
steady-state values.

To describe the key features of the NMR model, we divide the evolution into
five stages with the respective time intervals $[t_\mathrm{O},t_\mathrm{A}]$, 
$[t_\mathrm{A},t_\mathrm{B}]$, $[t_\mathrm{B},t_\mathrm{C}]$, 
$[t_\mathrm{C},t_\mathrm{D}]$, and $[t_\mathrm{D},t_\mathrm{E}]$, as shown in
Fig.~\ref{fg:X6}(a). In the earliest stage $[t_\mathrm{O},t_\mathrm{A}]$, 
$[X]_{t}$ increases due to internal failures but $[X]_{t}$ is insufficient 
to cause external failures. The behavior is similar to that in the MR model, 
but the duration is shorter and the rise in $[X]_{t}$ is steeper in the NMR 
model. The reason is that the memory effect in NMR model allows the recovery 
of $X$-type nodes to take place only after $\tau_{1}$ steps, while the 
recovery occurs probabilistically in the MR model. In the narrow time window 
of $[t_\mathrm{A},t_\mathrm{B}]$, $[X]_{t}$ attains a level high enough to 
trigger the onset of many external failures. As a result, the failed nodes 
constitute the majority in the system and $[A]_{t}$ decreases sharply, 
giving rise to the sharp increase in $[Y]_{t}$. The $Y$-type nodes recover 
deterministically after $\tau_{2}$ ($\tau_{2} < \tau_{1}$) into active nodes. 
In the period $[t_\mathrm{B},t_\mathrm{C}]$, the recovery of $Y$-type nodes 
refuels the system with active nodes that can participate in two paths: more 
internal and external failures. For $t_\mathrm{C} < \tau_{1}$, the existing 
$X$-type nodes have yet to recover and $[X]_{t}$ continues to increase but 
at a slower pace due to the external failure path, while $[A]_{t}$ reduces 
slightly.

In the time window $[t_\mathrm{C},t_\mathrm{D}]$, the initial internally 
failed nodes begin to recover as $t_\mathrm{C} > \tau_{1}$, in addition to 
the recovery of the $Y$-type nodes. The $A$-type nodes due to recovery will 
be more likely to become $Y$-type as the failed nodes remain the majority 
(due to the parameter setting $\beta_{2} > \beta_{1}$ in this example). 
This leads to the observed increase in $[Y]_{t}$ and decrease in $[X]_{t}$ 
in the time interval $[t_\mathrm{C},t_\mathrm{D}]$. In the final stage 
$[t_\mathrm{D},t_\mathrm{E}]$, $[X]_{t}$ stops decreasing because the 
recovery of $X$-type nodes at the time $t\gtrsim t_\mathrm{D}$ is due to 
those failed internally at $t \gtrsim t_\mathrm{B}$ for which the number 
was small. However, the recovery of $Y$-type nodes at a shorter time
scale supplies fresh active nodes. The fraction of failed nodes
$[X]_{t} + [Y]_{t}$ is so high, i.e., approaching the high-failure state, that
the dynamics lead to a higher steady value of $[Y]$ than $[X]$ in long time.
For time well beyond $t_\mathrm{D}$, both $[X]$ and $[Y]$ become steady.

Figure~\ref{fg:X6}(b) shows the phase diagram of the NMR model analogous to
Fig.~\ref{fg:X5}(b) for the MR model, with $\mu_1=0.01$ and $\mu_2=1.0$.
The results of the PA analysis (solid curve) are in better agreement with
the simulation results than those obtained from the mean-field theory
(dashed curve). The difference in dynamics in the NMR model also alters the
dependence of $\beta_\mathrm{c}$ to sustain a high-failure state on $[X]_{0}$.
Carrying out the same analysis as for the MR model (see Supplementary Figure 1 
in Supplementary Note 2 for details), we get the relationship 
$\beta_\mathrm{c}([X]_{0})$ for attaining a high-failure state for a given 
initial condition, as shown in Fig.~\ref{fg:X6}(c). The pair approximation, 
again, gives more accurate prediction than that from the mean-field theory.

The result in Fig.~\ref{fg:X6}(c) demonstrates the striking effect of
non-Markovian type of recovery with memory on the failure propagation
dynamics, which is in stark contrast to the ideal case of Markovian
process as exemplified in Fig.~\ref{fg:X5}(c). In particular, as
the fraction $[X]_{0}$ of initially failed nodes is increased from a near
zero value to one, the value of $\beta_\mathrm{c}$ begins to decrease 
continuously and smoothly until it reaches a minimum, at which 
$\beta_\mathrm{c}$ increases relatively more rapidly to a high value of 
about 0.006 for $[X]_{0} \approx 0.3$. For $[X]_{0} > 0.3$, the value of 
$\beta_\mathrm{c}$ remains approximately constant at 0.006. Comparing 
Fig.~\ref{fg:X6}(c) with Fig.~\ref{fg:X5}(c), we see two major, 
characteristic differences. Firstly, the behavior of an abrupt decrease in 
the Markovian case is replaced by a gradual process in the non-Markovian 
case, essentially converting a first-order like process to a second-order 
one. Secondly and more importantly, $\beta_\mathrm{c}$ recovers from its 
minimum value and maintains at a high value regardless of the value of 
$[X]_{0}$ insofar as it exceeds about $30\%$. This means that, the system 
can maintain its degree of resilience even when the initial fraction of 
failed nodes reaches $100\%$! This contrasts squarely the behavior in the 
Markovian case where the system resilience is reduced dramatically even 
when only about $40\%$ of the nodes failed initially. In this sense we say 
that a non-Markovian type of memory effect makes the network system more 
resilient against failure propagation.

While the behavior in Fig.~\ref{fg:X6}(c) is counterintuitive, a heuristic
reason is as follows. For an initial state with many initial $X$-type nodes,
the few remaining nodes will switch from being active to the $Y$-type and
back. All the initial $X$-type nodes will have to wait for the time period
$\tau_{1}$ to recover. At that time, the system becomes one with only a few
failed nodes - effectively equivalent to one with small $[X]_{0}$ value and
requiring a larger $\beta_\mathrm{c}$ value to evolve into the high-failure 
state. In a range of small $[X]_{0}$, a smaller $\beta_\mathrm{c}$ can 
already cause more active nodes to become $Y$-type, helping maintain the 
system in a high-failure state as described for Fig.~\ref{fg:X6}(a). 
Theoretical support for the behavior is provided by the PA analysis and 
mean-field theory, as shown in Fig.~\ref{fg:X6}(c).

\begin{figure*}
\includegraphics[width=0.8\linewidth]{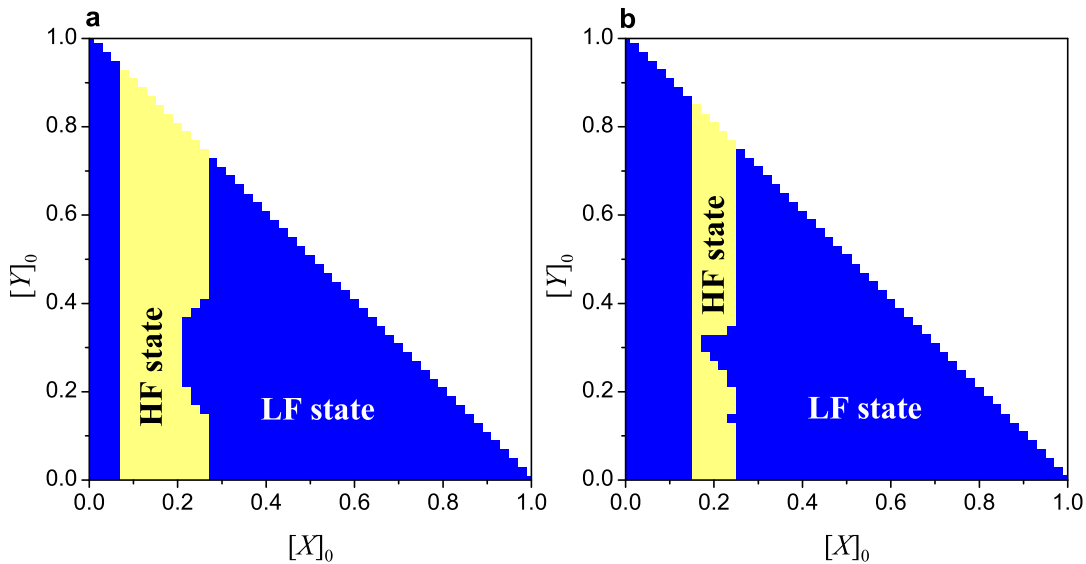}
\caption{ {\bf Basin structure of NMR model}. On the $[X]_0$-$[Y]_0$
plane, basin structure from (a) mean-field theory and (b) simulations. The
colors indicate the nature of the steady states given the initial conditions
$([X]_{0},[Y]_{0})$. The parameter setting is the same as that in
Fig.~\ref{fg:X3} for the MR model. The simulation results are obtained by
averaging over ten statistical realizations.}
\label{fg:X7}
\end{figure*}

In addition to the different time evolution in the MR and NMR models,
there are also cases where the same initial conditions $[X]_{0}$, $[Y]_{0}$,
and $[A]_{0}$ would lead to different final states. Figure~\ref{fg:X7} shows
the final states starting from any $[X]_{0}$ and $[Y]_{0}$ in the
$[X]_0$-$[Y]_0$ plane (the basin structure), with $\beta_1=0.004$,
$\beta_2=2.0$, $\mu_1=0.01$, and $\mu_2=1.0$. The results from the mean-field
theory [Fig.~\ref{fg:X7}(a)] and direct simulations [Fig.~\ref{fg:X7}(b)] show
essentially the same features. (Results from the initial-condition setting
$[X]_0\neq{0}$ and $[Y]_0=0.0$ are presented in Supplementary Figure 2 of 
Supplementary Note 2.) It is useful to contrast the final states of the MR 
and NMR models. From Fig.~\ref{fg:X3}, an initial state, e.g., 
$[X]_0$=$[Y]_0=0.5$, will evolve into a high-failure state in the MR model, 
but it will end up in a low-failure state in the NMR model. This means that, 
the NMR process can make the system more resilient to failures. (More 
examples can be found in Supplementary Figure 3 of Supplementary Note 2 
where different steady states from the two models are presented.)

\subsection{MR and NMR dynamics on heterogeneous networks}

\begin{figure} [ht!]
\includegraphics[width=0.8\linewidth]{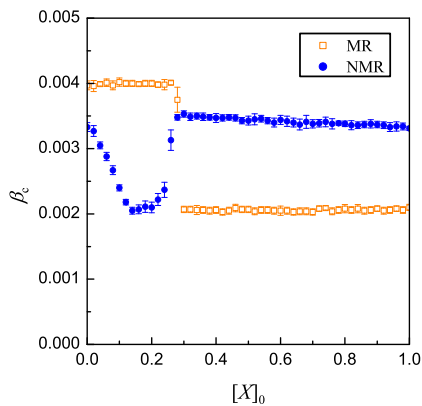}
\caption{ {\bf MR and NMR dynamics on heterogeneous networks}. The networks
are scale-free with $N=30,000$ nodes, degree exponent $\gamma = 3$,
$k_\mathrm{min}=6$ and $k_\mathrm{max}=173$. Shown is the dependence of
$\beta_\mathrm{c}$ for reaching a high-failure state on the initial value of
$[X]_{0}$, with $[Y]_{0}=0$, in the MR and NMR models for $\beta_2=1.9$,
$\mu_1=0.01$, $\mu_2=1$, $\tau_1=100$, and $\tau_2=1$. The results
qualitatively consistent with those in Fig.~\ref{fg:X5}(c) for MR dynamics
and in Fig.~\ref{fg:X6}(c) for NMR dynamics on random regular networks.}
\label{fg:hetero}
\end{figure}

So far our analysis and simulations have been carried out for MR and
NMR dynamics on random regular networks. We find that altering the network
structure causes little change in the qualitative results.
For example, we have carried out simulations on scale-free networks of
size $N=3\times 10^4$ with degree range $[k_\mathrm{min},\sqrt{N}]$ and 
degree distribution $P(k) \sim k^{-\gamma}$. Figure~\ref{fg:hetero} shows 
the results of $\beta_\mathrm{c}$ versus $[X]_{0}$
for the MR and NMR dynamics for networks with $\gamma=3$. Because of the
heterogeneity in the nodal degree distribution, the threshold on external
failure is given in terms of the fraction one-half of the failed neighbors.

Comparing results with Fig.~\ref{fg:X5}(c) for MR dynamics and
Fig.~\ref{fg:X6}(c) for NMR dynamics in random regular networks, we see that
the key features are similar when the underlying random regular networks are
replaced by scale-free networks.
We have also carried out numerical simulations on four additional types of
synthetic and empirical networks: (a) networks with degree-degree correlation,
(b) networks with a community structure, (c) empirical arenas-email network,
and (d) empirical friendship-hamster network, with results presented in
Supplementary Notes 4 and 5 for the former and latter two cases, respectively.
These results, together with Fig.~\ref{fg:hetero}, suggest that, for
heterogeneous networks, a non-Markovian process tends to enhance the network
resilience against large-scale failures.

\section*{Discussion} \label{sec:discussion}

The intrinsic memory effect associated with non-Markovian processes makes it 
challenging to analyze the underlying network dynamics, new and surprising 
phenomena can arise. Most previous studies treated Markovian processes through 
either a mean-field type of theory~\cite{Bottcher2017,Bottcher2017a} or an 
effective degree approach~\cite{Valdez2016}. For non-Markovian processes, the 
mean-field approximation can still be applied~\cite{majdandzic2014spontaneous,
Podobnik2015,Podobnik2015a,Majdandzic2016}, but it is necessary to invoke
a higher-order theory such as the PA analysis. Our work presents such an
example in the context of failure propagation in complex networks.

Our study has demonstrated that, in both models, the network can evolve
into a low-failure or a high-failure state, with the latter corresponding to
the undesired state of large-scale failure. Both the mean-field and PA theories
are capable of predicting the dynamical behaviors of failure propagation, and
the performances of the theories are gauged by simulation results, revealing
that the more laborious pair approximation gives results in better
quantitative agreement with the numerics. Our systematic computations
on different complex networks and two types of theoretical analyses have
uncovered a striking phenomenon: the non-Markovian memory effect in the
nodal recovery can counter-intuitively make the network more resilient
against large-scale failures.

Our finding also calls for the incorporation
of non-Markovian type of memory factors into the design of communication,
computer, and infrastructure networks in various engineering disciplines.
We hope our work will stimulate interest in examining and exploiting
non-Markovian processes in various network dynamical processes.
We have carried out a systematic study of the effects of Markovian versus
non-Markovian recovery on network synchronization using the paradigmatic
Kuramoto network model, with the main finding that non-Markovian recovery
makes the network more resilient against large-scale breakdown of
synchronization (Supplementary Note 6).

\section*{Methods}

{\bf A. Mean-field theory for MR dynamics}.
Let $[A]_t$, $[X]_t$ and $[Y]_t$ be the fractions of $A$-type, $X$-type and
$Y$-type nodes in the system at time $t$, respectively. A hierarchical set
of dynamical equations for the MR model can be constructed
to include increasingly longer spatial correlation. The equations for the
evolution of the fractions of different types of nodes are:
\begin{equation} \label{eq:mfr_x}
\frac{d[X]_{t}}{dt}=\beta_{1}[A]_{t}-\mu_{1}[X]_{t}
\end{equation}
and
\begin{equation} \label{eq:mfr_y}
\frac{d[Y]_{t}}{dt}=\beta_{2}E_t[A]_{t}-\mu_{2}[Y]_{t},
\end{equation}
where the first term in each equation gives the supply to $[X]$ ($[Y]$) due
to internal (external) failures and the second term represents the drop in
$[X]$ ($[Y]$) due to recovery. Note that, because of the relation
\begin{equation}
[A]_t=1-[X]_t-[Y]_t \equiv 1 - [I]_{t},
\end{equation}	
an equation for $[A]_{t}$ is unnecessary. The quantity $E_{t}$ is the
probability of an $A$-type node having $j\leq{m}$ neighbors of $A$-type
nodes at time $t$ and thus the node will be infected at the rate $\beta_{2}$.

In general, the quantity $E_{t}$ involves the correlation between two
neighboring nodes. To connect Eqs.~(\ref{eq:mfr_x}) and (\ref{eq:mfr_y})
so as to retain the simplicity of a single-node theory, we use the
approximation
\begin{equation} \label{eq:mfr_mf_e}
E_t([I])=\sum_{j=0}^{m}{C_{k}^{k-j}}{([I]_{t})}^{k-j}(1-[I]_{t})^{j},
\end{equation}
where $C_{k}^{k-j}=k!/(j!(k-j)!)$. Equations~(\ref{eq:mfr_x}-\ref{eq:mfr_mf_e})
form a set of equations, from which the fractions of different types of nodes
can be solved. This is the simplest single-site mean-field approximation for
the MR dynamics that ignores any spatial correlation. Despite its simplicity,
it is capable of revealing the key features in the stationary state, in which
Eqs.~(\ref{eq:mfr_x}) and (\ref{eq:mfr_y}) require the fraction of failed
nodes $[I]$ to satisfy
\begin{equation} \label{eq:selfI}
[I] = 1 - \frac{1}{ (\beta_2/\mu_2) E_t ([I]) + (\beta_1/\mu_1) + 1},
\end{equation}
which can be solved for $[I]$ self-consistently with Eq.~(\ref{eq:mfr_mf_e}).
Equation~(\ref{eq:selfI}) implies that $[I]$ depends only on the ratios
$\beta_{1}/\mu_{1}$ and $\beta_{2}/\mu_{2}$ within the mean-field
approximation, and so are the other fractions $[A]$, $[X]$, and $[Y]$.

{\bf B. Effect of nodal correlation: pairwise approximation for the MR model}.
Our PA based analysis begins by defining $[UV]_t$ as the fractions of $UV$
type of links in the system at time $t$, where $U, V \in \{A,X,Y\}$. A
connection that stems out from a node can be classified by a type. For
example, for a node with the current state being $A$-type, each link that it
carries can be classified into the $AA$, $AX$, or $AY$ type, depending on the
state of the node at the other end of the link. Taking into account every
link from every node, we have that the fractions of links satisfy
\begin{equation}
\sum_{U,V \in \{A,X,Y\}} [UV]_{t} =1,
\end{equation}
with $[UV]_t=[VU]_t$ for $U \neq V$.

\begin{figure}
\includegraphics[width=\linewidth]{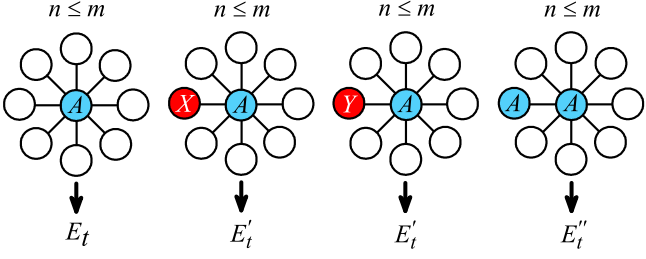}
\caption{ {\bf Schematic diagram for the PA analysis of the MR model}.
The quantities are $E_t$, $E_t^{'}$ and
$E_t^{''}$. The blue (red) color indicates a node in the active (failure)
state. Open circles are nodes that may take on $A$, $X$, or $Y$ state.
In the PA equations, $E_{t}$ is the probability of an $A$-type node having
$n\leq{m}$ neighbors of $A$-type nodes at time $t$, $E^{'}_{t}$ is the
probability of an $A$-type node having $n\leq{m}$ $A$-type neighbors among
its $(k-1)$ neighbors given that one neighbor is inactive, and $E{''}_{t}$
is the probability of an $A$-type node having $n\leq{m-1}$ $A$-type neighbors
among its $(k-1)$ neighbors given that one neighbor is active.}
\label{fig:e_diagram}
\end{figure}

In general, the equations of single-node quantities, e.g., Eq.~(\ref{eq:mfr_y}),
necessarily involve quantities of more extensive spatial correlation because
the interplay between the failure of a node and the states of its neighboring
nodes. Since $[AI]_{t}/[A]_{t} = ([AX]_t+[AY]_t)/[A]_{t}$ is the probability
of an $A$-type node having an inactive node regardless of the types of the
neighbors, the probability that there are exactly $j$ neighbors of $A$-type
and $(k-j)$ inactive neighbors of either $X$ or $Y$ type is
\begin{equation}
{C_{k}^{k-j}}\left(\frac{[AI]_{t}}{[A]_{t}}\right)^{k-j}\left(1-\frac{[AI]_{t}}{[A]_{t}}\right)^{j},
\end{equation}
where $k$ is the degree of the node. The quantity $E_{t}$ in
Eq.~(\ref{eq:mfr_y}), as schematically depicted in Fig.~\ref{fig:e_diagram}(a),
is thus given by
\begin{equation} \label{eq:E_t}
E_{t}=\sum_{j=0}^{m}{C_{k}^{k-j}}\left(\frac{[AI]_{t}}{[A]_{t}}\right)^{k-j}
\left(1-\frac{[AI]_{t}}{[A]_{t}}\right)^{j},
\end{equation}
which indicates explicitly that the dynamics of single-node quantities are
governed by the two-node quantity $[AI]_{t}$. This is reminiscence of the
BBGKY (Bogoliubov-Born-Green-Kirkwood-Yvon) hierarchy of equations for the
distribution functions in a system consisting of a large number of interacting
particles in statistical physics~\cite{Harris:book}. Only under the
approximation $[AI]_{t} \approx [A]_{t} [I]_{t}$ (so that the two-node
correlation can be neglected) will the resulting equation be
Eq.~(\ref{eq:mfr_mf_e}) - a set of single-node mean-field equations.

To proceed, we derive the dynamical equations for $[UV]_{t}$ that will in
general involve more extensive spatial correlation. For example, a link of
the type $AA$ would evolve into a different type depending on the neighborhoods
of the two nodes, effectively a small cluster of nodes. To develop a
manageable approximation, we retain the two-node correlation and decouple
any longer spatial correlation in terms of one-node and two-node functions.
This is the idea behind PA for obtaining a closed set of equations. In
particular, the dynamical equations for $[AX]_{t}$ and $[AA]_{t}$ are
\begin{equation} \label{eq:mfr_ax}
\begin{split}
\frac{d[AX]_{t}}{dt}=&\mu_{1}[XX]_{t}+\mu_{2}[YX]_{t}+\beta_{1}[AA]_{t}\\
&-\mu_{1}[AX]_{t}-(\beta_{1}+\beta_{2}E^{'}_{t})[AX]_{t}\\
\end{split}
\end{equation}
and
\begin{equation} \label{eq:mfr_aa}
\begin{split}
\frac{d[AA]_{t}}{dt}=&2\mu_{1}[AX]_{t}+2\mu_{2}[AY]_{t}\\
&-2(\beta_{1}+\beta_{2}E^{''}_{t})[AA]_{t},\\
\end{split}
\end{equation}
where
\begin{equation} \label{eq:Et'}
{E^{'}_{t}}=\sum_{j=0}^{m}{C_{k-1}^{k-1-j}}
\left(\frac{[AI]_{t}}{[A]_{t}}\right)^{k-1-j}
\left(1-\frac{[AI]_{t}}{[A]_{t}}\right)^{j}
\end{equation}
is the probability of an $A$-type node having $j\leq{m}$ $A$-type neighbors
among its $(k-1)$ neighbors, given that one neighbor is inactive, and
\begin{equation} \label{eq:Et''}
{E^{''}_{t}}=\sum_{j=0}^{m-1}{C_{k-1}^{k-1-j}}
\left(\frac{[AI]_{t}}{[A]_{t}}\right)^{k-1-j}
\left(1-\frac{[AI]_{t}}{[A]_{t}}\right)^{j}
\end{equation}
is the probability of an $A$-type node having $j\leq{m-1}$ $A$-type neighbors
among its $(k-1)$ neighbors, given that one neighbor is active.
Figure~\ref{fig:e_diagram} illustrates the meanings of $E_{t}$, $E_{t}^{'}$,
and $E_{t}^{''}$ schematically. The terms in Eqs.~(\ref{eq:mfr_ax}) and
(\ref{eq:mfr_aa}) account for how the recovery and failure processes affect
the fractions of $AX$-type and $AA$-type links. The complete set of
dynamical equations is listed in Supplementary Note 1, which can be
solved iteratively to yield the temporal variations on the type of nodes and
the type of links given an initial condition. The steady-state quantities can
be obtained through a sufficiently large number of iterations.

{\bf C. Pairwise approximation theory for the NMR model}.
Specifically, we let $[U^l]_t$ be the
fraction of nodes of type $U$ at time $t$, which became type $U$ from some
other type only $l$ time steps ago, and $[U^{l_1}V^{l_2}]_t$
be the fraction of links of the UV type when the corresponding node(s)
associated with a link became that of the labeled type $l_{1}$ and $l_{2}$
time steps ago. The time evolution of the fraction of $X$-type nodes in the
NMR model is given by
\begin{equation} \label{eq:nmfr_pa_x}
[X^{l}]_{t+\Delta{t}}=\left\{
\begin{aligned}
&\beta_{1}\Delta{t}[A]_{t}, & & {{l}\in[0,\Delta{t});}\\
&[X^{l-\Delta{t}}]_{t}, & & {{l}\in[\Delta{t},\tau_{1}];}\\
&0, & & {l\in(\tau_{1},\infty).}
\end{aligned}
\right.
\end{equation}
The first line in Eq.~(\ref{eq:nmfr_pa_x}) gives the new supply due to
internal failure of $A$-type nodes in the time duration $[t,t+\Delta{t})$.
The second line accounts for the nodes which were inactive for a duration
$l-\Delta t$ at time $t$ but have not reached the time for recovery at
time $t+\Delta t$. The third line states that all $X$-type nodes that came
to existence $\tau_{1}$ earlier have been recovered. Similarly, the time
evolution of the fraction of $Y$-type nodes is given by
\begin{equation} \label{eq:nmfr_pa_y}
[Y^{l}]_{t+\Delta{t}}=\left\{
\begin{aligned}
&\beta_{2}\Delta{t}E_t[A]_{t}, & & {{l}\in[0,\Delta{t});}\\
&[Y^{l-\Delta{t}}]_{t}, & & {{l}\in[\Delta{t},\tau_{2}]};\\
&0, & & {l\in(\tau_{2},\infty),}
\end{aligned}
\right.
\end{equation}
where $E_{t}$ is defined in Eq.~(\ref{eq:E_t}) and $[AI]_t=[AX]_t+[AY]_t$.
The fractions of $X$-type and $Y$-type nodes, regardless of how long they
have been in the corresponding state, are given by
$[X]_t=\sum_{l=0}^{\tau_1}[X^l]_t$ and $[Y]_t=\sum_{l=0}^{\tau_2}[Y^l]_t$,
respectively. The fraction of active nodes follows from $[A]_t=1-[X]_t-[Y]_t$.

To develop a PA analysis for failure propagation dynamics with NMR, we
construct the equations for the time evolution of $UV$-types of links and
retain spatial correlation up to two neighboring nodes. Our derivation of
the counterparts of Eqs.~(\ref{eq:nmfr_pa_x}) and (\ref{eq:nmfr_pa_y}) in
the MR case suggests the necessity to examine the history of the inactive
nodes(s) associated with a link. For example, the time evolution of the
links in $[AX^{l}]_{t}$ is governed by
\begin{equation} \label{eq:nmfr_pa_ax}
[AX^{l}]_{t+\Delta{t}}=\left\{
\begin{aligned}
&\beta_{1}\Delta{t}[AA]_{t}+\beta_{1}\Delta{t}([X^{\tau_{1}}A]_{t} \\
&+[Y^{\tau_{2}}A]_{t}), & & {{l}\in[0,\Delta{t});}\\
&[X^{\tau_1}X^{l-\Delta{t}}]_{t}+[Y^{\tau_2}X^{l-\Delta{t}}]_{t} \\
&+(1-\beta_{1}\Delta{t}-\beta_{2}\Delta{t}{E^{'}_{t}}) \\
&\times [AX^{l-\Delta{t}}]_{t}, & & {{l}\in[\Delta{t},\tau_{1}]};\\
&0, & & {l\in(\tau_{1},\infty),}
\end{aligned}
\right.
\end{equation}
where $E_{t}'$ is defined in Eq.~(\ref{eq:Et'}). The first line represents
the new supply to $AX$-type of links due to an internal failure in one of
the active nodes associated with a link of the $AA$-type, and an internal
failure together with a recovery of an inactive node in a link of the
$XA$- and $YA$-types. The second line includes the supply to $AX$$^{l}$-type
links due to recoveries from $XX$ and $YX$ types as well as the links of
$AX$$^{l-\Delta{t}}$ type that became $AX$$^{l}$ type in the recent
duration $\Delta t$. The last line comes from the fact that an $X$-type node
must recover after a time $\tau_{1}$ since it became inactive. The fraction
of links of $AX$-type, regardless of how long the node in the link has taken
in the $X$-type, is given by $[AX]_t=\sum_{l=0}^{\tau_1}[AX^l]_t$. We thus
have that the fraction of $AA$-type of links evolves in time as
\begin{eqnarray} \label{eq:nmfr_pa_aa}
[AA]_{t+\Delta{t}}&=&2(1-\beta_{1}\Delta{t}-\beta_{2}\Delta{t}{E^{'}_{t}})(
[AX^{\tau_{1}}]_{t}+[AY^{\tau_{2}}]_{t}) \nonumber \\
& & +[X^{\tau_1}X^{\tau_1}]_{t}+[Y^{\tau_2}Y^{\tau_2}]_{t}+2[X^{\tau_1}Y^{\tau_2}]_{t}  \nonumber \\
& & +(1-2\beta_{1}\Delta{t}-2\beta_{2}\Delta{t}{E^{''}_{t}})[AA]_{t},
\end{eqnarray}
where $E^{''}_{t}$ is defined in Eq.~(\ref{eq:Et''}). Equations for other
types of links can also be constructed (Supplementary Note 1).
Equations~(\ref{eq:nmfr_pa_ax}) and (\ref{eq:nmfr_pa_aa}) are analogous to
Eqs.~(\ref{eq:mfr_ax}) and (\ref{eq:mfr_aa}) in the MR model. The number of
equations is determined by the divisions of $\tau_{1}$ and $\tau_{2}$ into the
small time steps $\Delta t$, which increases rapidly when $\Delta t$ is small
compared with the other time scales in the NMR dynamics.

A crude approximation analogous to the mean-field theory can be developed for
the NMR model by retaining only the fractions of nodes in the equations, which
can be done by decoupling the two-node quantities such as $[AI]_{t}$ by
$[AI]_t \approx [A]_t [I]_t$. The resulting equations governing the fractions
of different types of nodes become
\begin{equation} \label{eq:nmfr_mf_x}
[X]_{t+\Delta{t}}=\beta_{1}\Delta{t}[A]_{t}+[X]_{t}-[X^{\tau_1}]_{t},
\end{equation}
and
\begin{equation} \label{eq:nmfr_mf_y}
[Y]_{t+\Delta{t}}=\beta_{2}\Delta{t}E_t[A]_{t}+[Y]_{t}-[Y^{\tau_2}]_{t},
\end{equation}
where $E_{t}$ takes on the approximate form in Eq.~(\ref{eq:mfr_mf_e}).
Equations~(\ref{eq:nmfr_mf_x}), (\ref{eq:nmfr_mf_y}), and (\ref{eq:mfr_mf_e})
form a set of equations that can be solved to yield the fractions of different
types of nodes. The first two terms in Eqs.~(\ref{eq:nmfr_mf_x}) and
(\ref{eq:nmfr_mf_y}) correspond to the increase in inactive nodes due to
failure and due to those remaining inactive, and the last term corresponds to
recovery. The number of equations, again, depends on the choice of $\Delta t$.
This is the mean-field approximation for the NMR model that ignores any
spatial correlation.

\section*{Data Availability}

The source data underlying Figs. 2-7 and Supplementary Figs. 1-12 are available at https://github.com/zhlin2328/Codes-for-NCOMMS-19-1125220.

\section*{Code Availability}

C++ codes to reproduce the data in the main text and the Supplementary Information are available at https://github.com/zhlin2328/Codes-for-NCOMMS-19-1125220.


\section*{Reference}
\begin{thebibliography}{100}
\expandafter\ifx\csname url\endcsname\relax
  \def\url#1{\texttt{#1}}\fi
\expandafter\ifx\csname urlprefix\endcsname\relax\def\urlprefix{URL }\fi
\providecommand{\bibinfo}[2]{#2}
\providecommand{\eprint}[2][]{\url{#2}}

\bibitem{ML:2002}
\bibinfo{author}{Motter, A.~E.} \& \bibinfo{author}{Lai, Y.-C.}
\newblock \bibinfo{title}{Cascade-based attacks on complex networks}.
\newblock \emph{\bibinfo{journal}{Phys. Rev. E}} \textbf{\bibinfo{volume}{66}},
  \bibinfo{pages}{065102} (\bibinfo{year}{2002}).

\bibitem{ZPL:2004}
\bibinfo{author}{Zhao, L.}, \bibinfo{author}{Park, K.} \& \bibinfo{author}{Lai,
  Y.-C.}
\newblock \bibinfo{title}{Attack vulnerability of scale-free networks due to
  cascading breakdown}.
\newblock \emph{\bibinfo{journal}{Phys. Rev.E}} \textbf{\bibinfo{volume}{70}},
  \bibinfo{pages}{035101(R)} (\bibinfo{year}{2004}).

\bibitem{ZPLY:2005}
\bibinfo{author}{Zhao, L.}, \bibinfo{author}{Park, K.}, \bibinfo{author}{Lai,
  Y.-C.} \& \bibinfo{author}{Ye, N.}
\newblock \bibinfo{title}{Tolerance of scale-free networks against
  attack-induced cascades}.
\newblock \emph{\bibinfo{journal}{Phys. Rev.E}} \textbf{\bibinfo{volume}{72}},
  \bibinfo{pages}{025104(R)} (\bibinfo{year}{2005}).

\bibitem{GC:2007}
\bibinfo{author}{Galstyan, A.} \& \bibinfo{author}{Cohen, P.}
\newblock \bibinfo{title}{Cascading dynamics in modular networks}.
\newblock \emph{\bibinfo{journal}{Phys. Rev. E}} \textbf{\bibinfo{volume}{75}},
  \bibinfo{pages}{036109} (\bibinfo{year}{2007}).

\bibitem{Bialek:2007}
\bibinfo{author}{Bialek, J.~W.}
\newblock \bibinfo{title}{Why has it happened again? {Comparison between the
  UCTE} blackout in 2006 and the blackouts of 2003}.
\newblock In \emph{\bibinfo{booktitle}{Power Tech 2007 IEEE Lausanne}},
  \bibinfo{pages}{51--56} (\bibinfo{publisher}{IEEE}, \bibinfo{year}{2007}).

\bibitem{dobson2007complex}
\bibinfo{author}{Dobson, I.}, \bibinfo{author}{Carreras, B.~A.},
  \bibinfo{author}{Lynch, V.~E.} \& \bibinfo{author}{Newman, D.~E.}
\newblock \bibinfo{title}{Complex systems analysis of series of blackouts:
  Cascading failure, critical points, and self-organization}.
\newblock \emph{\bibinfo{journal}{Chaos}} \textbf{\bibinfo{volume}{17}},
  \bibinfo{pages}{026103} (\bibinfo{year}{2007}).

\bibitem{Gleeson:2008}
\bibinfo{author}{Gleeson, J.~P.}
\newblock \bibinfo{title}{Cascades on correlated and modular random networks}.
\newblock \emph{\bibinfo{journal}{Phys. Rev. E}} \textbf{\bibinfo{volume}{77}},
  \bibinfo{pages}{046117} (\bibinfo{year}{2008}).

\bibitem{rosato2008modelling}
\bibinfo{author}{Rosato, V.} \emph{et~al.}
\newblock \bibinfo{title}{Modelling interdependent infrastructures using
  interacting dynamical models}.
\newblock \emph{\bibinfo{journal}{Int. J. Crit. Infrastruc.}}
  \textbf{\bibinfo{volume}{4}}, \bibinfo{pages}{63--79} (\bibinfo{year}{2008}).

\bibitem{HLC:2008}
\bibinfo{author}{Huang, L.}, \bibinfo{author}{Lai, Y.-C.} \&
  \bibinfo{author}{Chen, G.}
\newblock \bibinfo{title}{Understanding and preventing cascading breakdown in
  complex clustered networks}.
\newblock \emph{\bibinfo{journal}{Phys. Rev. E}} \textbf{\bibinfo{volume}{78}},
  \bibinfo{pages}{036116} (\bibinfo{year}{2008}).

\bibitem{SBPBH:2008}
\bibinfo{author}{Simonsen, I.}, \bibinfo{author}{Buzna, L.},
  \bibinfo{author}{Peters, K.}, \bibinfo{author}{Bornholdt, S.} \&
  \bibinfo{author}{Helbing, D.}
\newblock \bibinfo{title}{Transient dynamics increasing network vulnerability
  to cascading failures}.
\newblock \emph{\bibinfo{journal}{Phys. Rev. Lett.}}
  \textbf{\bibinfo{volume}{100}}, \bibinfo{pages}{218701}
  (\bibinfo{year}{2008}).

\bibitem{YWLC:2009}
\bibinfo{author}{Yang, R.}, \bibinfo{author}{Wang, W.-X.},
  \bibinfo{author}{Lai, Y.-C.} \& \bibinfo{author}{Chen, G.}
\newblock \bibinfo{title}{Optimal weighting scheme for suppressing cascades and
  traffic congestion in complex networks}.
\newblock \emph{\bibinfo{journal}{Phys. Rev. E}} \textbf{\bibinfo{volume}{79}},
  \bibinfo{pages}{026112} (\bibinfo{year}{2009}).

\bibitem{takayasu2010econophysics}
\bibinfo{author}{Takayasu, M.}, \bibinfo{author}{Watanabe, T.} \&
  \bibinfo{author}{Takayasu, H.}
\newblock \emph{\bibinfo{title}{Econophysics Approaches to Large-Scale Business
  Data and Financial Crisis: Proceedings of Tokyo Tech-Hitotsubashi
  Interdisciplinary Conference and APFA7}} (\bibinfo{publisher}{Springer
  Science \& Business Media}, \bibinfo{year}{2010}).

\bibitem{HL:2011}
\bibinfo{author}{Huang, L.} \& \bibinfo{author}{Lai, Y.-C.}
\newblock \bibinfo{title}{Cascading dynamics in complex quantum networks}.
\newblock \emph{\bibinfo{journal}{Chaos}} \textbf{\bibinfo{volume}{21}},
  \bibinfo{pages}{025107} (\bibinfo{year}{2011}).

\bibitem{WLA:2011}
\bibinfo{author}{Wang, W.}, \bibinfo{author}{Lai, Y.-C.} \&
  \bibinfo{author}{Armbruster, D.}
\newblock \bibinfo{title}{Cascading failures and the emergence of cooperation
  in evolutionary-game based models of social and economical networks}.
\newblock \emph{\bibinfo{journal}{Chaos}} \textbf{\bibinfo{volume}{21}},
  \bibinfo{pages}{033112} (\bibinfo{year}{2011}).

\bibitem{LWLW:2012}
\bibinfo{author}{Liu, R.-R.}, \bibinfo{author}{Wang, W.-X.},
  \bibinfo{author}{Lai, Y.-C.} \& \bibinfo{author}{Wang, B.-H.}
\newblock \bibinfo{title}{Cascading dynamics on random networks: Crossover in
  phase transition}.
\newblock \emph{\bibinfo{journal}{Phys. Rev. E}} \textbf{\bibinfo{volume}{85}},
  \bibinfo{pages}{026110} (\bibinfo{year}{2012}).

\bibitem{li2015percolation}
\bibinfo{author}{Li, D.} \emph{et~al.}
\newblock \bibinfo{title}{Percolation transition in dynamical traffic network
  with evolving critical bottlenecks}.
\newblock \emph{\bibinfo{journal}{Proc. Nat. Acad. Sci. (USA)}}
  \textbf{\bibinfo{volume}{112}}, \bibinfo{pages}{669--672}
  (\bibinfo{year}{2015}).

\bibitem{parshani2011critical}
\bibinfo{author}{Parshani, R.}, \bibinfo{author}{Buldyrev, S.~V.} \&
  \bibinfo{author}{Havlin, S.}
\newblock \bibinfo{title}{Critical effect of dependency groups on the function
  of networks}.
\newblock \emph{\bibinfo{journal}{Proc. Nat. Acad. Sci. (USA)}}
  \textbf{\bibinfo{volume}{108}}, \bibinfo{pages}{1007--1010}
  (\bibinfo{year}{2011}).

\bibitem{watts2002simple}
\bibinfo{author}{Watts, D.~J.}
\newblock \bibinfo{title}{A simple model of global cascades on random
  networks}.
\newblock \emph{\bibinfo{journal}{Proc. Nat. Acad. Sci. (USA)}}
  \textbf{\bibinfo{volume}{99}}, \bibinfo{pages}{5766--5771}
  (\bibinfo{year}{2002}).

\bibitem{dodds2004universal}
\bibinfo{author}{Dodds, P.~S.} \& \bibinfo{author}{Watts, D.~J.}
\newblock \bibinfo{title}{Universal behavior in a generalized model of
  contagion}.
\newblock \emph{\bibinfo{journal}{Phys. Rev. Lett.}}
  \textbf{\bibinfo{volume}{92}}, \bibinfo{pages}{218701}
  (\bibinfo{year}{2004}).

\bibitem{simonsen2008transient}
\bibinfo{author}{Simonsen, I.}, \bibinfo{author}{Buzna, L.},
  \bibinfo{author}{Peters, K.}, \bibinfo{author}{Bornholdt, S.} \&
  \bibinfo{author}{Helbing, D.}
\newblock \bibinfo{title}{Transient dynamics increasing network vulnerability
  to cascading failures}.
\newblock \emph{\bibinfo{journal}{Phys. Rev. Lett.}}
  \textbf{\bibinfo{volume}{100}}, \bibinfo{pages}{218701}
  (\bibinfo{year}{2008}).

\bibitem{buldyrev2010catastrophic}
\bibinfo{author}{Buldyrev, S.~V.}, \bibinfo{author}{Parshani, R.},
  \bibinfo{author}{Paul, G.}, \bibinfo{author}{Stanley, H.~E.} \&
  \bibinfo{author}{Havlin, S.}
\newblock \bibinfo{title}{Catastrophic cascade of failures in interdependent
  networks}.
\newblock \emph{\bibinfo{journal}{Nature}} \textbf{\bibinfo{volume}{464}},
  \bibinfo{pages}{1025} (\bibinfo{year}{2010}).

\bibitem{ganin2017resilience}
\bibinfo{author}{Ganin, A.~A.} \emph{et~al.}
\newblock \bibinfo{title}{Resilience and efficiency in transportation
  networks}.
\newblock \emph{\bibinfo{journal}{Sci. Adv.}} \textbf{\bibinfo{volume}{3}},
  \bibinfo{pages}{e1701079} (\bibinfo{year}{2017}).

\bibitem{nudo2013recovery}
\bibinfo{author}{Nudo, R.~J.}
\newblock \bibinfo{title}{Recovery after brain injury: mechanisms and
  principles}.
\newblock \emph{\bibinfo{journal}{Front. Human Neurosci.}}
  \textbf{\bibinfo{volume}{7}}, \bibinfo{pages}{887} (\bibinfo{year}{2013}).

\bibitem{shang2015impact}
\bibinfo{author}{Shang, Y.}
\newblock \bibinfo{title}{Impact of self-healing capability on network
  robustness}.
\newblock \emph{\bibinfo{journal}{Phys. Rev. E}} \textbf{\bibinfo{volume}{91}},
  \bibinfo{pages}{042804} (\bibinfo{year}{2015}).

\bibitem{hu2016recovery}
\bibinfo{author}{Hu, F.}, \bibinfo{author}{Yeung, C.~H.},
  \bibinfo{author}{Yang, S.}, \bibinfo{author}{Wang, W.} \&
  \bibinfo{author}{Zeng, A.}
\newblock \bibinfo{title}{Recovery of infrastructure networks after localised
  attacks}.
\newblock \emph{\bibinfo{journal}{Sci. Rep.}} \textbf{\bibinfo{volume}{6}},
  \bibinfo{pages}{24522} (\bibinfo{year}{2016}).

\bibitem{white2001autonomic}
\bibinfo{author}{White, S.~R.} \emph{et~al.}
\newblock \bibinfo{title}{Autonomic healing of polymer composites}.
\newblock \emph{\bibinfo{journal}{Nature}} \textbf{\bibinfo{volume}{409}},
  \bibinfo{pages}{794} (\bibinfo{year}{2001}).

\bibitem{toohey2007self}
\bibinfo{author}{Toohey, K.~S.}, \bibinfo{author}{Sottos, N.~R.},
  \bibinfo{author}{Lewis, J.~A.}, \bibinfo{author}{Moore, J.~S.} \&
  \bibinfo{author}{White, S.~R.}
\newblock \bibinfo{title}{Self-healing materials with microvascular networks}.
\newblock \emph{\bibinfo{journal}{Nat. Mater.}} \textbf{\bibinfo{volume}{6}},
  \bibinfo{pages}{581} (\bibinfo{year}{2007}).

\bibitem{desmurget2007contrasting}
\bibinfo{author}{Desmurget, M.}, \bibinfo{author}{Bonnetblanc, F.} \&
  \bibinfo{author}{Duffau, H.}
\newblock \bibinfo{title}{Contrasting acute and slow-growing lesions: a new
  door to brain plasticity}.
\newblock \emph{\bibinfo{journal}{Brain}} \textbf{\bibinfo{volume}{130}},
  \bibinfo{pages}{898--914} (\bibinfo{year}{2007}).

\bibitem{majdandzic2014spontaneous}
\bibinfo{author}{Majdandzic, A.} \emph{et~al.}
\newblock \bibinfo{title}{Spontaneous recovery in dynamical networks}.
\newblock \emph{\bibinfo{journal}{Nat. Phys.}} \textbf{\bibinfo{volume}{10}},
  \bibinfo{pages}{34} (\bibinfo{year}{2014}).

\bibitem{podobnik2014network}
\bibinfo{author}{Podobnik, B.} \emph{et~al.}
\newblock \bibinfo{title}{Network risk and forecasting power in phase-flipping
  dynamical networks}.
\newblock \emph{\bibinfo{journal}{Phys. Rev. E}} \textbf{\bibinfo{volume}{89}},
  \bibinfo{pages}{042807} (\bibinfo{year}{2014}).

\bibitem{Podobnik2015}
\bibinfo{author}{Podobnik, B.} \emph{et~al.}
\newblock \bibinfo{title}{Predicting the lifetime of dynamic networks
  experiencing persistent random attacks}.
\newblock \emph{\bibinfo{journal}{Sci. Rep.}} \textbf{\bibinfo{volume}{5}},
  \bibinfo{pages}{14286} (\bibinfo{year}{2015}).

\bibitem{Podobnik2015a}
\bibinfo{author}{Podobnik, B.} \emph{et~al.}
\newblock \bibinfo{title}{The cost of attack in competing networks}.
\newblock \emph{\bibinfo{journal}{J. Roy. Soc. Interface}}
  \textbf{\bibinfo{volume}{12}}, \bibinfo{pages}{20150770}
  (\bibinfo{year}{2015}).

\bibitem{Majdandzic2016}
\bibinfo{author}{Majdandzic, A.} \emph{et~al.}
\newblock \bibinfo{title}{{Multiple tipping points and optimal repairing in
  interacting networks.}}
\newblock \emph{\bibinfo{journal}{Nat. Commun.}} \textbf{\bibinfo{volume}{7}},
  \bibinfo{pages}{10850} (\bibinfo{year}{2016}).

\bibitem{national2012disaster}
\bibinfo{author}{Council, N.~R.} \emph{et~al.}
\newblock \bibinfo{title}{Disaster resilience: A national imperative, the
  national academies}.
\newblock \emph{\bibinfo{journal}{Press, Washington DC}}
  (\bibinfo{year}{2012}).

\bibitem{gao2016universal}
\bibinfo{author}{Gao, J.}, \bibinfo{author}{Barzel, B.} \&
  \bibinfo{author}{Barab{\'a}si, A.-L.}
\newblock \bibinfo{title}{Universal resilience patterns in complex networks}.
\newblock \emph{\bibinfo{journal}{Nature}} \textbf{\bibinfo{volume}{530}},
  \bibinfo{pages}{307--312} (\bibinfo{year}{2016}).

\bibitem{ganin2016operational}
\bibinfo{author}{Ganin, A.~A.} \emph{et~al.}
\newblock \bibinfo{title}{Operational resilience: concepts, design and
  analysis}.
\newblock \emph{\bibinfo{journal}{Sci. Rep.}} \textbf{\bibinfo{volume}{6}},
  \bibinfo{pages}{1--12} (\bibinfo{year}{2016}).

\bibitem{linkov2019science}
\bibinfo{author}{Linkov, I.} \& \bibinfo{author}{Trump, B.~D.}
\newblock \emph{\bibinfo{title}{The Science and Practice of Resilience}}
  (\bibinfo{publisher}{Springer}, \bibinfo{year}{2019}).

\bibitem{pastor2015epidemic}
\bibinfo{author}{Pastor-Satorras, R.}, \bibinfo{author}{Castellano, C.},
  \bibinfo{author}{Van~Mieghem, P.} \& \bibinfo{author}{Vespignani, A.}
\newblock \bibinfo{title}{Epidemic processes in complex networks}.
\newblock \emph{\bibinfo{journal}{Rev. Mod. Phys.}}
  \textbf{\bibinfo{volume}{87}}, \bibinfo{pages}{925} (\bibinfo{year}{2015}).

\bibitem{wang2017unification}
\bibinfo{author}{Wang, W.}, \bibinfo{author}{Tang, M.},
  \bibinfo{author}{Stanley, H.~E.} \& \bibinfo{author}{Braunstein, L.~A.}
\newblock \bibinfo{title}{Unification of theoretical approaches for epidemic
  spreading on complex networks}.
\newblock \emph{\bibinfo{journal}{Rep. Prog. Phys.}}
  \textbf{\bibinfo{volume}{80}}, \bibinfo{pages}{036603}
  (\bibinfo{year}{2017}).

\bibitem{de2018fundamentals}
\bibinfo{author}{de~Arruda, G.~F.}, \bibinfo{author}{Rodrigues, F.~A.} \&
  \bibinfo{author}{Moreno, Y.}
\newblock \bibinfo{title}{Fundamentals of spreading processes in single and
  multilayer complex networks}.
\newblock \emph{\bibinfo{journal}{Phys. Rep.}} \textbf{\bibinfo{volume}{756}},
  \bibinfo{pages}{1--60} (\bibinfo{year}{2018}).

\bibitem{Barabasi:2005}
\bibinfo{author}{Barabasi, A.-L.}
\newblock \bibinfo{title}{The origin of bursts and heavy tails in human
  dynamics}.
\newblock \emph{\bibinfo{journal}{Nature}} \textbf{\bibinfo{volume}{435}},
  \bibinfo{pages}{207} (\bibinfo{year}{2005}).

\bibitem{GHB:2008}
\bibinfo{author}{Gonz\'{a}lez, M.~C.}, \bibinfo{author}{Hidalgo, C.~A.} \&
  \bibinfo{author}{Barab\'{a}si, A.~L.}
\newblock \bibinfo{title}{Understanding individual human mobility patterns}.
\newblock \emph{\bibinfo{journal}{Nature}} \textbf{\bibinfo{volume}{453}},
  \bibinfo{pages}{779--782} (\bibinfo{year}{2008}).

\bibitem{SGMB:2012}
\bibinfo{author}{Simini, F.}, \bibinfo{author}{Gonz\'{a}lez, M.~C.},
  \bibinfo{author}{Maritan, A.} \& \bibinfo{author}{Barab\'{a}si, A.~L.}
\newblock \bibinfo{title}{A universal model for mobility and migration
  patterns}.
\newblock \emph{\bibinfo{journal}{Nature}} \textbf{\bibinfo{volume}{484}},
  \bibinfo{pages}{96--100} (\bibinfo{year}{2012}).

\bibitem{ZYZZHL:2013}
\bibinfo{author}{Zhao, Z.-D.} \emph{et~al.}
\newblock \bibinfo{title}{Emergence of scaling in human-interest dynamics}.
\newblock \emph{\bibinfo{journal}{Sci. Rep.}} \textbf{\bibinfo{volume}{3}},
  \bibinfo{pages}{3472} (\bibinfo{year}{2013}).

\bibitem{ZHHLL:2014}
\bibinfo{author}{Zhao, Z.-D.}, \bibinfo{author}{Huang, Z.-G.},
  \bibinfo{author}{Huang, L.}, \bibinfo{author}{Liu, H.} \&
  \bibinfo{author}{Lai, Y.-C.}
\newblock \bibinfo{title}{Scaling and correlation of human movements in cyber
  and physical spaces}.
\newblock \emph{\bibinfo{journal}{Phys. Rev. E}} \textbf{\bibinfo{volume}{90}},
  \bibinfo{pages}{050802(R)} (\bibinfo{year}{2014}).

\bibitem{PSRPGB:2015}
\bibinfo{author}{Pappalardo, L.} \emph{et~al.}
\newblock \bibinfo{title}{Returners and explorers dichotomy in human mobility}.
\newblock \emph{\bibinfo{journal}{Nat. Commun.}} \textbf{\bibinfo{volume}{6}},
  \bibinfo{pages}{8166} (\bibinfo{year}{2015}).

\bibitem{ZZYWL:2016}
\bibinfo{author}{Zhao, Y.-M.}, \bibinfo{author}{Zeng, A.},
  \bibinfo{author}{Yan, X.-Y.}, \bibinfo{author}{Wang, W.-X.} \&
  \bibinfo{author}{Lai, Y.-C.}
\newblock \bibinfo{title}{Unified underpinning of human mobility in the real
  world and cyberspace}.
\newblock \emph{\bibinfo{journal}{New J. Phys.}} \textbf{\bibinfo{volume}{18}},
  \bibinfo{pages}{053025} (\bibinfo{year}{2016}).

\bibitem{YWGL:2017}
\bibinfo{author}{Yan, X.-Y.}, \bibinfo{author}{Wang, W.-X.},
  \bibinfo{author}{Gao, Z.-Y.} \& \bibinfo{author}{Lai, Y.-C.}
\newblock \bibinfo{title}{Universal model of individual and population mobility
  on diverse spatial scales}.
\newblock \emph{\bibinfo{journal}{Nat. Commun.}} \textbf{\bibinfo{volume}{8}},
  \bibinfo{pages}{1639} (\bibinfo{year}{2017}).

\bibitem{bratsun2005delay}
\bibinfo{author}{Bratsun, D.}, \bibinfo{author}{Volfson, D.},
  \bibinfo{author}{Tsimring, L.~S.} \& \bibinfo{author}{Hasty, J.}
\newblock \bibinfo{title}{Delay-induced stochastic oscillations in gene
  regulation}.
\newblock \emph{\bibinfo{journal}{Proc. Nat. Acad. Sci. (USA)}}
  \textbf{\bibinfo{volume}{102}}, \bibinfo{pages}{14593--14598}
  (\bibinfo{year}{2005}).

\bibitem{scalas2006waiting}
\bibinfo{author}{Scalas, E.}, \bibinfo{author}{Kaizoji, T.},
  \bibinfo{author}{Kirchler, M.}, \bibinfo{author}{Huber, J.} \&
  \bibinfo{author}{Tedeschi, A.}
\newblock \bibinfo{title}{Waiting times between orders and trades in
  double-auction markets}.
\newblock \emph{\bibinfo{journal}{Physica A}} \textbf{\bibinfo{volume}{366}},
  \bibinfo{pages}{463--471} (\bibinfo{year}{2006}).

\bibitem{VRLB:2007}
\bibinfo{author}{Vazquez, A.}, \bibinfo{author}{Racz, B.},
  \bibinfo{author}{Lukacs, A.} \& \bibinfo{author}{Barabasi, A.-L.}
\newblock \bibinfo{title}{Impact of {non-Poissonian} activity patterns on
  spreading processes}.
\newblock \emph{\bibinfo{journal}{Phys. Rev. Lett.}}
  \textbf{\bibinfo{volume}{98}}, \bibinfo{pages}{158702}
  (\bibinfo{year}{2007}).

\bibitem{IM:2009}
\bibinfo{author}{Iribarren, J.~L.} \& \bibinfo{author}{Moro, E.}
\newblock \bibinfo{title}{Impact of human activity patterns on the dynamics of
  information diffusion}.
\newblock \emph{\bibinfo{journal}{Phys. Rev. Lett.}}
  \textbf{\bibinfo{volume}{103}}, \bibinfo{pages}{038702}
  (\bibinfo{year}{2009}).

\bibitem{MB:2013}
\bibinfo{author}{Van~Mieghem, P.} \& \bibinfo{author}{Van~de Bovenkamp, R.}
\newblock \bibinfo{title}{{Non-Markovian} infection spread dramatically alters
  the susceptible-infected-susceptible epidemic threshold in networks}.
\newblock \emph{\bibinfo{journal}{Phys. Rev. Lett.}}
  \textbf{\bibinfo{volume}{110}}, \bibinfo{pages}{108701}
  (\bibinfo{year}{2013}).

\bibitem{JPKK:2014}
\bibinfo{author}{Jo, H.-H.}, \bibinfo{author}{Perotti, J.~I.},
  \bibinfo{author}{Kaski, K.} \& \bibinfo{author}{Kert{\'e}sz, J.}
\newblock \bibinfo{title}{Analytically solvable model of spreading dynamics
  with {non-Poissonian} processes}.
\newblock \emph{\bibinfo{journal}{Phys. Rev. X}} \textbf{\bibinfo{volume}{4}},
  \bibinfo{pages}{011041} (\bibinfo{year}{2014}).

\bibitem{KRV:2015}
\bibinfo{author}{Kiss, I.~Z.}, \bibinfo{author}{R{\"o}st, G.} \&
  \bibinfo{author}{Vizi, Z.}
\newblock \bibinfo{title}{Generalization of pairwise models to {non-Markovian}
  epidemics on networks}.
\newblock \emph{\bibinfo{journal}{Phys. Rev. Lett.}}
  \textbf{\bibinfo{volume}{115}}, \bibinfo{pages}{078701}
  (\bibinfo{year}{2015}).

\bibitem{SGB:2017}
\bibinfo{author}{Starnini, M.}, \bibinfo{author}{Gleeson, J.~P.} \&
  \bibinfo{author}{Bogu{\~n}{\'a}, M.}
\newblock \bibinfo{title}{Equivalence between {non-Markovian and Markovian}
  dynamics in epidemic spreading processes}.
\newblock \emph{\bibinfo{journal}{Phys. Rev. Lett.}}
  \textbf{\bibinfo{volume}{118}}, \bibinfo{pages}{128301}
  (\bibinfo{year}{2017}).

\bibitem{SMBK:2018}
\bibinfo{author}{Sherborne, N.}, \bibinfo{author}{Miller, J.},
  \bibinfo{author}{Blyuss, K.} \& \bibinfo{author}{Kiss, I.}
\newblock \bibinfo{title}{Mean-field models for {non-Markovian} epidemics on
  networks}.
\newblock \emph{\bibinfo{journal}{J. Math. Biol.}}
  \textbf{\bibinfo{volume}{76}}, \bibinfo{pages}{755--558}
  (\bibinfo{year}{2018}).

\bibitem{FCTL:2019}
\bibinfo{author}{Feng, M.}, \bibinfo{author}{Cai, S.-M.},
  \bibinfo{author}{Tang, M.} \& \bibinfo{author}{Lai, Y.-C.}
\newblock \bibinfo{title}{Equivalence and its invalidation between
  {non-Markovian and Markovian} spreading dynamics on complex networks}.
\newblock \emph{\bibinfo{journal}{Nat. Commun.}} \textbf{\bibinfo{volume}{10}},
  \bibinfo{pages}{3748} (\bibinfo{year}{2019}).

\bibitem{Valdez2016}
\bibinfo{author}{Valdez, L.~D.}, \bibinfo{author}{{Di Muro}, M.~A.} \&
  \bibinfo{author}{Braunstein, L.~A.}
\newblock \bibinfo{title}{{Failure-recovery model with competition between
  failures in complex networks: A dynamical approach}}.
\newblock \emph{\bibinfo{journal}{J. Stat. Mech. Theo. Exp.}}
  \textbf{\bibinfo{volume}{2016}} (\bibinfo{year}{2016}).

\bibitem{Bottcher2017}
\bibinfo{author}{B{\"{o}}ttcher, L.}, \bibinfo{author}{Nagler, J.} \&
  \bibinfo{author}{Herrmann, H.~J.}
\newblock \bibinfo{title}{Critical behaviors in contagion dynamics}.
\newblock \emph{\bibinfo{journal}{Phys. Rev. Lett.}}
  \textbf{\bibinfo{volume}{118}}, \bibinfo{pages}{1--5} (\bibinfo{year}{2017}).

\bibitem{Bottcher2017a}
\bibinfo{author}{B{\"{o}}ttcher, L.}, \bibinfo{author}{Lukovi{\'{c}}, M.},
  \bibinfo{author}{Nagler, J.}, \bibinfo{author}{Havlin, S.} \&
  \bibinfo{author}{Herrmann, H.~J.}
\newblock \bibinfo{title}{{Failure and recovery in dynamical networks}}.
\newblock \emph{\bibinfo{journal}{Sci. Rep.}} \textbf{\bibinfo{volume}{7}},
  \bibinfo{pages}{41729} (\bibinfo{year}{2017}).

\bibitem{keeling1997correlation}
\bibinfo{author}{Keeling, M.}, \bibinfo{author}{Rand, D.} \&
  \bibinfo{author}{Morris, A.}
\newblock \bibinfo{title}{Correlation models for childhood epidemics}.
\newblock \emph{\bibinfo{journal}{Proc. Royal Soc. London. Ser. B Biol. Sci.}}
  \textbf{\bibinfo{volume}{264}}, \bibinfo{pages}{1149--1156}
  (\bibinfo{year}{1997}).

\bibitem{ben1992mean}
\bibinfo{author}{ben Avraham, D.} \& \bibinfo{author}{K{\"o}hler, J.}
\newblock \bibinfo{title}{Mean-field (n, m)-cluster approximation for lattice
  models}.
\newblock \emph{\bibinfo{journal}{Phys. Rev. A}} \textbf{\bibinfo{volume}{45}},
  \bibinfo{pages}{8358} (\bibinfo{year}{1992}).

\bibitem{mata2013pair}
\bibinfo{author}{Mata, A.~S.} \& \bibinfo{author}{Ferreira, S.~C.}
\newblock \bibinfo{title}{Pair quenched mean-field theory for the
  susceptible-infected-susceptible model on complex networks}.
\newblock \emph{\bibinfo{journal}{EPL (Europhys. Lett.)}}
  \textbf{\bibinfo{volume}{103}}, \bibinfo{pages}{48003}
  (\bibinfo{year}{2013}).

\bibitem{gross2006epidemic}
\bibinfo{author}{Gross, T.}, \bibinfo{author}{D’Lima, C. J.~D.} \&
  \bibinfo{author}{Blasius, B.}
\newblock \bibinfo{title}{Epidemic dynamics on an adaptive network}.
\newblock \emph{\bibinfo{journal}{Phys. Rev. Lett.}}
  \textbf{\bibinfo{volume}{96}}, \bibinfo{pages}{208701}
  (\bibinfo{year}{2006}).

\bibitem{ji2013coevolve}
\bibinfo{author}{Ji, M.}, \bibinfo{author}{Xu, C.}, \bibinfo{author}{Choi,
  C.~W.} \& \bibinfo{author}{Hui, P.~M.}
\newblock \bibinfo{title}{Correlation and analytic approaches to co-evolving
  voter models}.
\newblock \emph{\bibinfo{journal}{New J. Phys.}} \textbf{\bibinfo{volume}{15}},
  \bibinfo{pages}{113024} (\bibinfo{year}{2013}).

\bibitem{zhang2013SG}
\bibinfo{author}{Zhang, W.}, \bibinfo{author}{Xu, C.} \& \bibinfo{author}{Hui,
  P.~M.}
\newblock \bibinfo{title}{Spatial structure enhanced cooperation in
  dissatisfied adaptive snowdrift game}.
\newblock \emph{\bibinfo{journal}{Euro. Phys. J. B}}
  \textbf{\bibinfo{volume}{86}}, \bibinfo{pages}{196} (\bibinfo{year}{2013}).

\bibitem{zhang2014coevolve}
\bibinfo{author}{Zhang, W.}, \bibinfo{author}{Li, Y.~S.}, \bibinfo{author}{Du,
  P.}, \bibinfo{author}{Xu, C.} \& \bibinfo{author}{Hui, P.~M.}
\newblock \bibinfo{title}{Phase transitions in a coevolving snowdraft game with
  costly rewiring}.
\newblock \emph{\bibinfo{journal}{Phys. Rev. E}} \textbf{\bibinfo{volume}{90}},
  \bibinfo{pages}{052819} (\bibinfo{year}{2014}).

\bibitem{choi2017coevolve}
\bibinfo{author}{Choi, C.~W.}, \bibinfo{author}{Xu, C.} \&
  \bibinfo{author}{Hui, P.~M.}
\newblock \bibinfo{title}{Adaptive cyclically dominating game on co-evolving
  networks: numerical and analytic reuslts}.
\newblock \emph{\bibinfo{journal}{Euro. Phys. J. B}}
  \textbf{\bibinfo{volume}{90}}, \bibinfo{pages}{190} (\bibinfo{year}{2017}).

\bibitem{Harris:book}
\bibinfo{author}{Harris, S.}
\newblock \emph{\bibinfo{title}{An Introduction to the Theory of the Boltzmann
  Equation}} (\bibinfo{publisher}{Courier Corporation}, \bibinfo{year}{2004}).

\end{thebibliography}

\section*{Acknowledgments}

The authors would like to thank Zhenhua Wang for helpful discussions. 
This work was supported by the National Natural Science Foundation of China
(Grant Nos. 11975099, 11575041, 11675056 and 11835003), the Natural Science
Foundation of Shanghai (Grant No. 18ZR1412200), and the Science and Technology
Commission of Shanghai Municipality (Grant No. 14DZ2260800). YCL would like
to acknowledge support from the Vannevar Bush Faculty Fellowship program
sponsored by the Basic Research Office of the Assistant Secretary of Defense
for Research and Engineering and funded by the Office of Naval Research
through Grant No.~N00014-16-1-2828.

\section*{Author Contributions}

Z.-H.L., M.T. and Z.H.L. designed research; Z.-H.L. performed research; 
Z.-H.L., M.F., M.T., Z.H.L., C.X. and P.M.H. contributed analytic tools; 
Z.-H.L., M.F., M.T., Z.H.L., C.X., P.M.H. and Y.-C.L. analyzed data; 
Z.-H.L., M.T., Z.H.L., P.M.H. and Y.-C.L. wrote the paper.

\section*{Competing Interests}

The authors declare no competing interests.

\section*{Correspondence}

To whom correspondence should be addressed. E-mail: tangminghan007@gmail.com; zhliu@phy.ecnu.edu.cn

\end{document}


\begin{center}
{\large Supplementary Information for}
\end{center}
\begin{center}
{\Large\bf Non-Markovian recovery makes complex networks more resilient against large scale failures}
\end{center}
\begin{center}
          Zhao-Hua Lin, Mi Feng, Ming Tang, Zonghua Liu, Chen Xu, Pak Ming Hui and Ying-Cheng Lai
\end{center}
\begin{center}
Corresponding author: Ming Tang (tangminghan007@gmail.com), Zonghua Liu(zhliu@phy.ecnu.edu.cn)
\end{center}

\renewcommand\figurename{Supplementary Figure}
\renewcommand\thefigure{\arabic{figure}}
\renewcommand\theequation{\arabic{equation}}

\tableofcontents

\newpage
\section{Supplementary Figures}

\begin{figure} [H]
	\includegraphics[width=\linewidth]{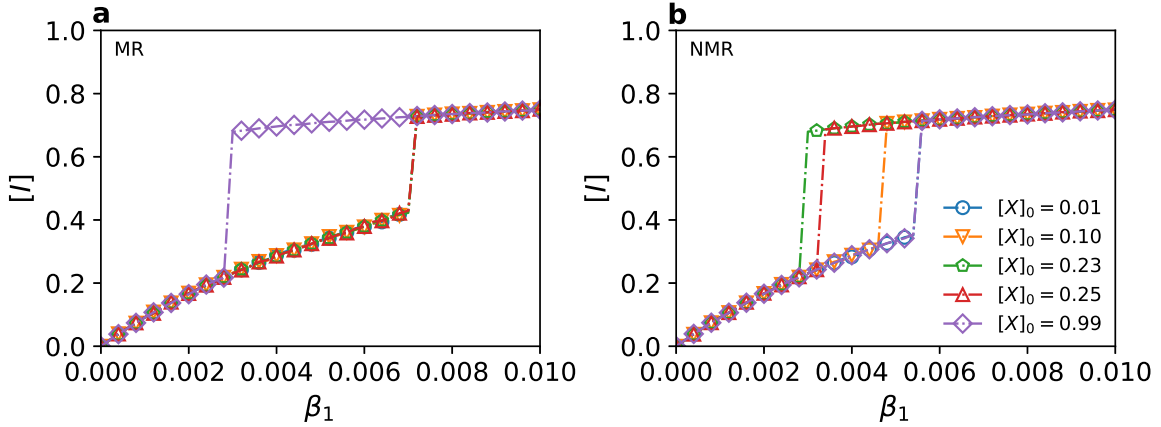}
	\caption{{\bf Effects of initial conditions on phase transition}. 
		The initial conditions are $[X]_{0} \neq 0$ and $[Y]_{0}=0$. The phase
		transition is with respect to a systematic increase in the value of
		parameter $\beta_{1}$. The results are obtained from the same initial
		conditions for different values of $\beta_1$ (for a fixed $\beta_{2}$ value).
		There is a critical value $\beta_\textrm{c}([X]_{0})$ beyond which the system
		approaches a high-failure state. (a,b) Results for MR and NMR models,
		respectively, where the blue circles, orange down triangles, green pentagons,
		red up triangles, purple diamonds represent the results from different initial
		fractions of failed nodes: $[X]_0=0.01, 0.1, 0.23, 0.25, 0.99$, respectively.
		Other parameters are $\beta_2=2$, $\mu_1=0.01$, $\mu_2=1$, $\tau_1=100$,
		$\tau_2=1$, and $m=15$. The network has a random regular structure with
		size $N=3\times 10^4$ and nodal degree $k=35$.}
	\label{fig:rho0_betaC}
\end{figure}

\begin{figure} [H]
	\includegraphics[width=\linewidth]{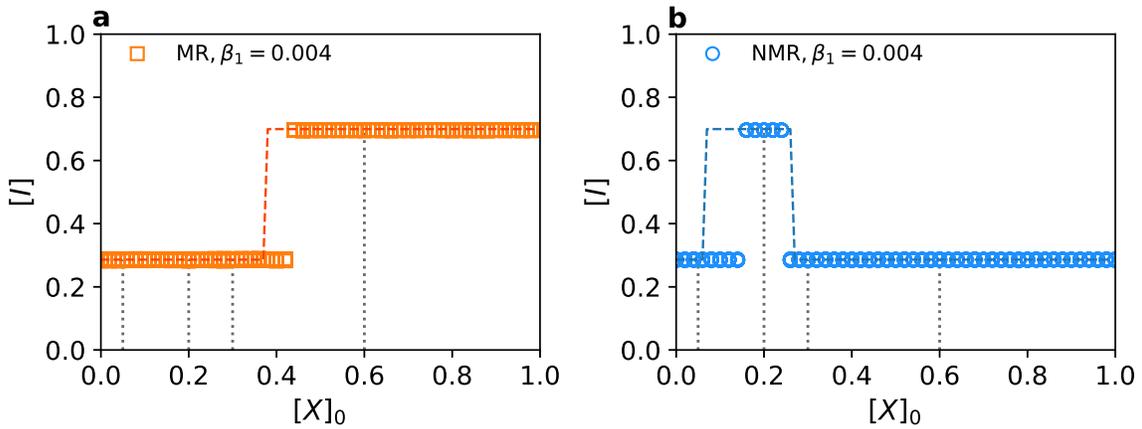}
	\caption{{\bf Effects of initial conditions on stationary solution}. 
		The initial conditions are $[X]_{0} \neq 0$ and $[Y]_{0}=0$. The stationary
		solution $[I]$ is obtained for $\beta_1=0.004$ for MR (a) and NMR (b) models.
		Orange squares and blue circles are simulation results for MR and NMR models,
		respectively. The dashed line represents the mean-field prediction. The gray
		dotted vertical lines correspond to $[X]_0=0.05, 0.2, 0.3, 0.6$, respectively.}
	\label{fig:rho0_rho}
\end{figure}

\begin{figure*} [ht!]
	\includegraphics[width=0.7\linewidth]{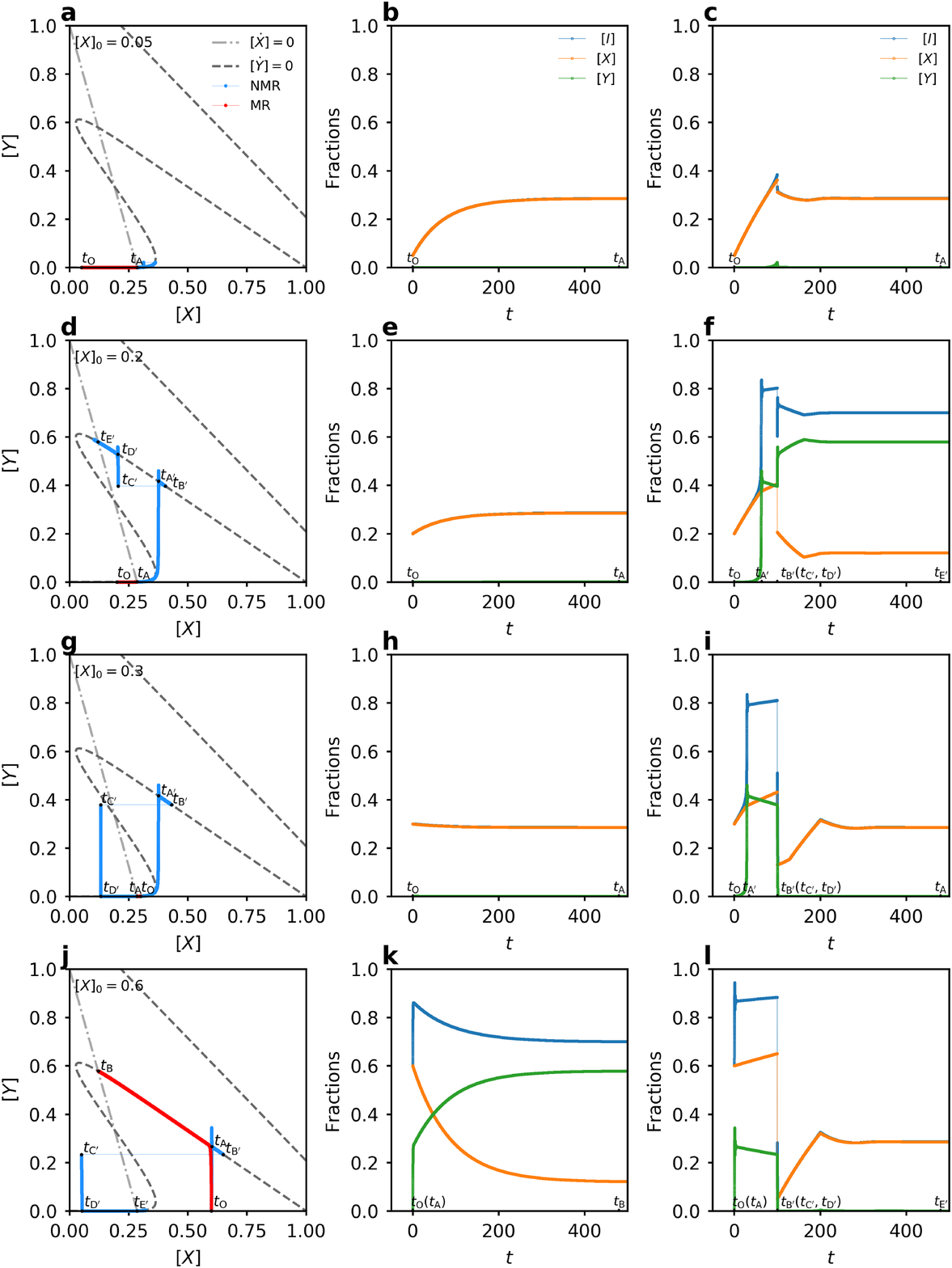}
	\caption{{\bf Trajectories and time evolution of fractions of $X$-type and $Y$-type nodes}. 
The initial conditions are $[X]_{0} \neq 0$ and $[Y]_{0}=0$.
		The results are obtained from the mean-field theory. For fixed $[X]_0=0.05$,
		(a) trajectories of $[X]$ and $[Y]$, (b,c) time evolution
		from the MR and NMR models, respectively. (d-f) The corresponding results
		for $[X]_0=0.2$. (g-i) The results for $[X]_0=0.3$. (j-l) The results for
		$[X]_0=0.6$. The solid blue and red lines in the first column are the
		results from the NMR and MR models, respectively. The light and dark gray
		dotted lines are the solutions of $[\dot{X}]=0$ and $[\dot{Y}]=0$ from the
		mean-field theory for the MR model, respectively, where their intersections
		give the steady-state solutions. The solid blue, orange and green lines in
		the second and third columns are the results of $[I]$, $[X]$ and $[Y]$ for
		the MR and NMR models, respectively. The evolution from $t_{\textrm{B}'}$
		to $t_{\textrm{C}'}$ and then to $t_{\textrm{D}'}$ is too fast to be distinguished in
		(f), (i) and (l). The evolution from $t_{\textrm{O}}$ to $t_{\textrm{A}}$ is also too short
		for it to be seen in (k) and (l). Other parameter values are $\beta_1=0.004$,
		$\beta_2=2$, $\mu_1=0.01$, $\mu_2=1$, $\tau_1=100$, $\tau_2=1$, and $m=15$.
		The network is the same as that in Supplementary Fig.~\ref{fig:rho0_betaC}.}
	\label{fig:trajectory_and evolution}
\end{figure*}

\begin{figure} [H]
	\includegraphics[width=\linewidth]{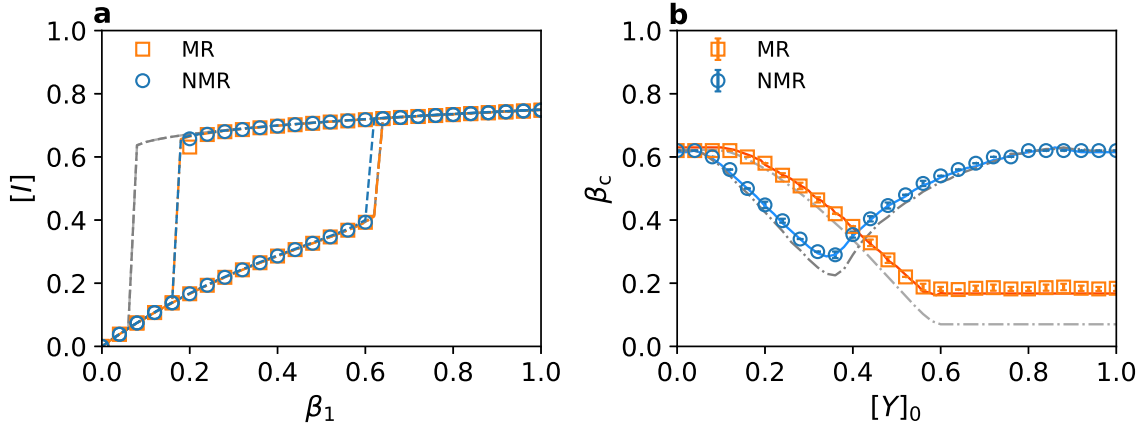}
	\caption{{\bf MR and NMR dynamics in situations where external recovery is slower than internal recovery}. 
(a) Dependence of $[I]$ on $\beta_1$ in the steady state for $\beta_2=0.1$,
		$\tau_1=1.0$ (corresponding to $\mu_1=1.0$),
		$\tau_2=20$ (corresponding to $\mu_2=0.05$), and $m=15$.
		Orange squares (blue circles) are simulation results for the MR (NMR)
		model. The results are averaged for two network configurations, each of ten
		realizations. The orange dot-dashed (blue dashed) line is calculated by the
		PA theory for the MR (NMR) model. The gray dot-dashed (dashed) line is the
		result from the	MF theory for the MR (NMR) model. (b) Dependence of $\beta_\textrm{c}$ on the initial value of $[Y]_{0}$, with $[X]_{0}=0$. The solid (dot-dashed)
		lines are obtained from the PA (MF) theory. The networks are RRNs
		with $N=30000$ and $k=35$.}
	\label{fig:F1}
\end{figure}

\begin{figure} [H]
	\includegraphics[width=\linewidth]{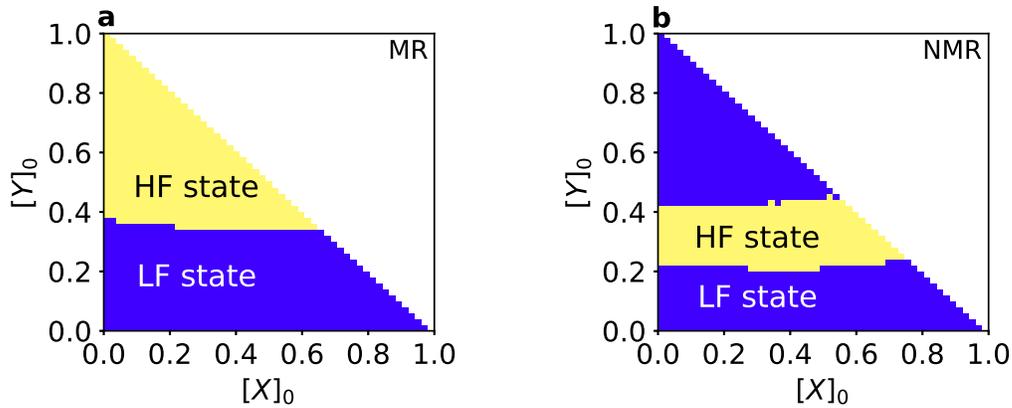}
	\caption{{\bf Basin structures}. On the $[X]_0$-$[Y]_0$ plane, basin structure for
		(a) MR and (b) NMR model for $\beta_1=0.4$, where the colors indicate the
		nature of the steady states from different initial conditions. Other
		parameters are the same as those in Supplementary Fig.~\ref{fig:F1}.}
	\label{fig:F2}
\end{figure}

\begin{figure} [H]
	\includegraphics[width=\linewidth]{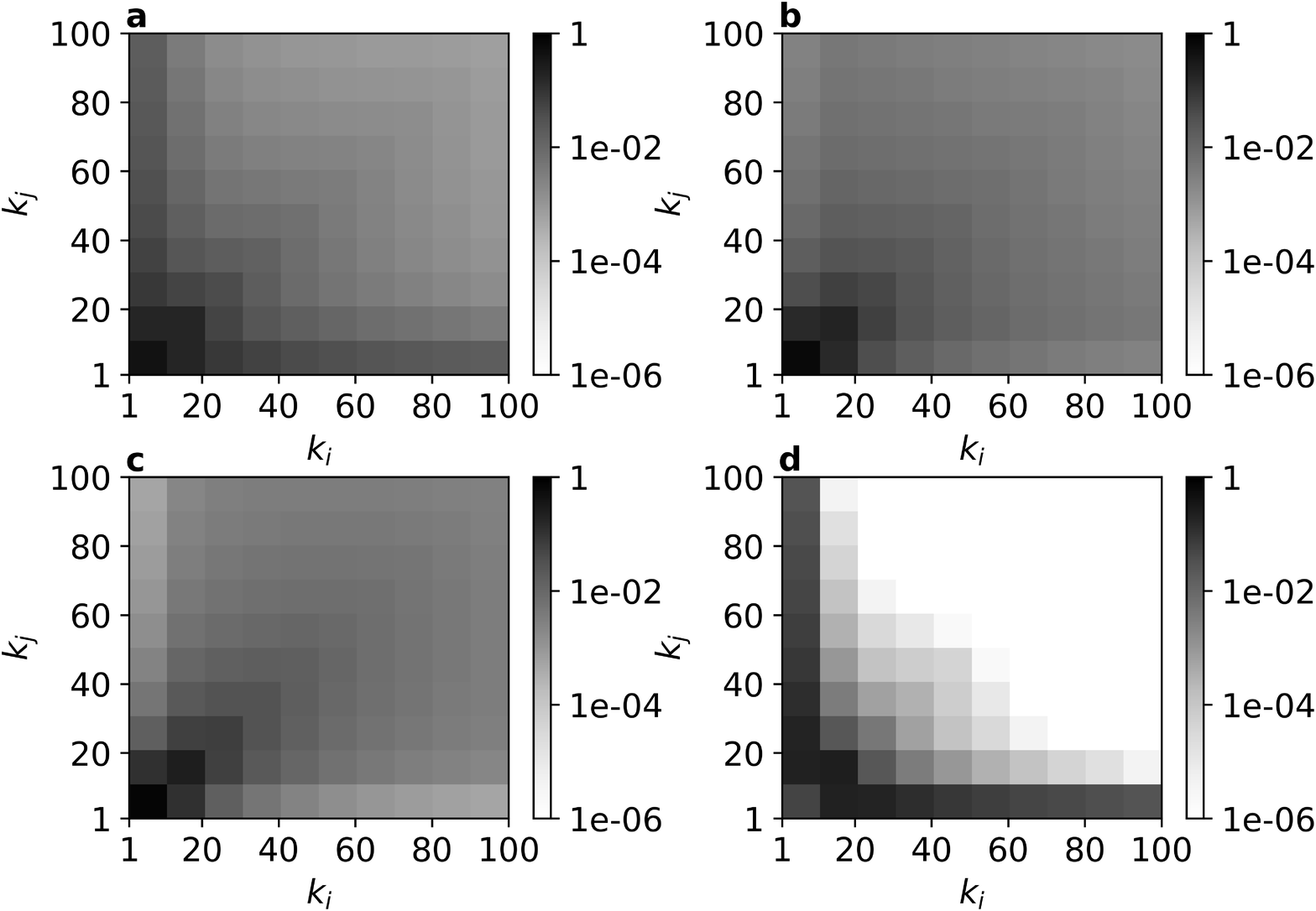}
	\caption{{\bf Visualizations of degree-degree correlation in a network}. Shading
		colors of the matrix correspond to the probabilities that a randomly chosen
		edge connects nodes $i$ of degree $k_i$ and $j$ of $k_j$, where the
		results are averaged with $100$ UCN realizations for (a) $r=0$,
		(b) $r=0.5$, (c) $r=0.7$ and (d) $r=-0.5$. Each grid cell represents the
		average result with the degree range of ten for visualization. Other network
		parameters are $N=10000$, $\gamma=2.5$, $k_\textrm{min}=5$, and $k_\textrm{max}=100$.}
	\label{fig:aij}
\end{figure}

\begin{figure} [H]
	\includegraphics[width=\linewidth]{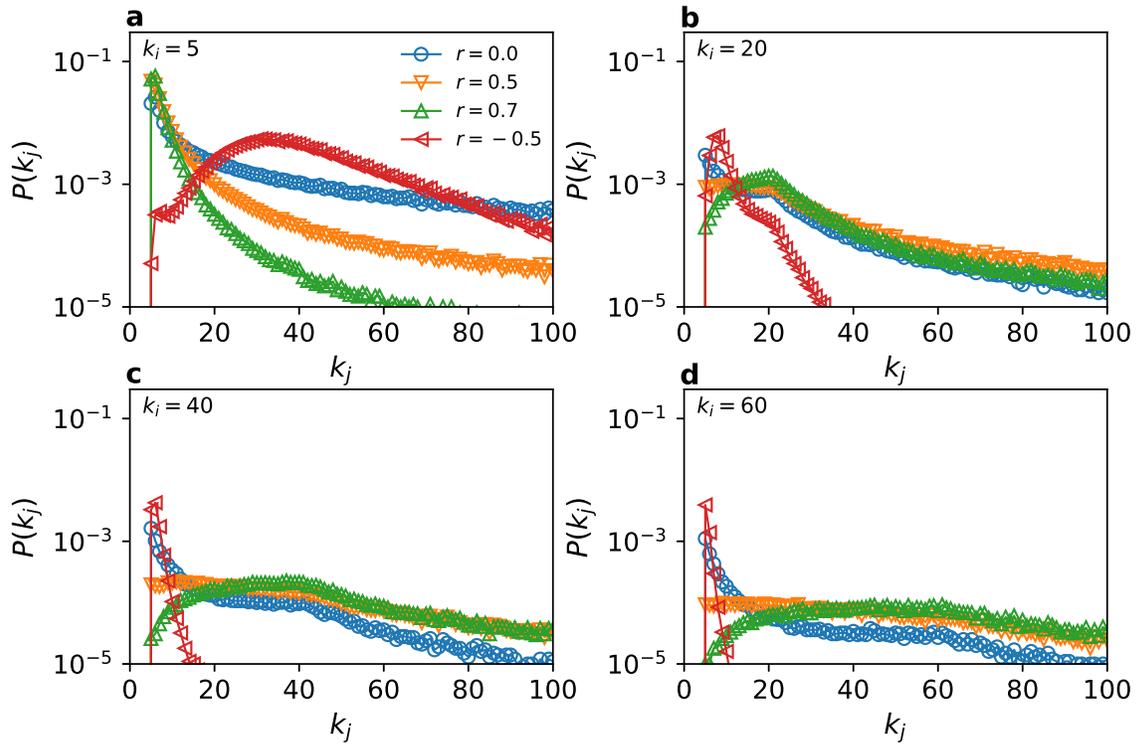}
	\caption{{\bf Probability distribution of a node of degree $k_i$ connecting to a node of degree $k_j$}. 
		(a-d): node $i$ of degree $k_i=5$, $k_i=20$, $k_i=40$ and $k_i=60$,
		respectively. Blue circles, orange down triangles, green up triangles and
		red left triangles are the results for different level of degree-degree correlation $r=0$, $r=0.5$, $r=0.7$ 
and $r=-0.5$ in Supplementary Figs.~\ref{fig:aij}, respectively.}
	\label{fig:dist_knn}
\end{figure}

\begin{figure} [H]
	\includegraphics[width=\linewidth]{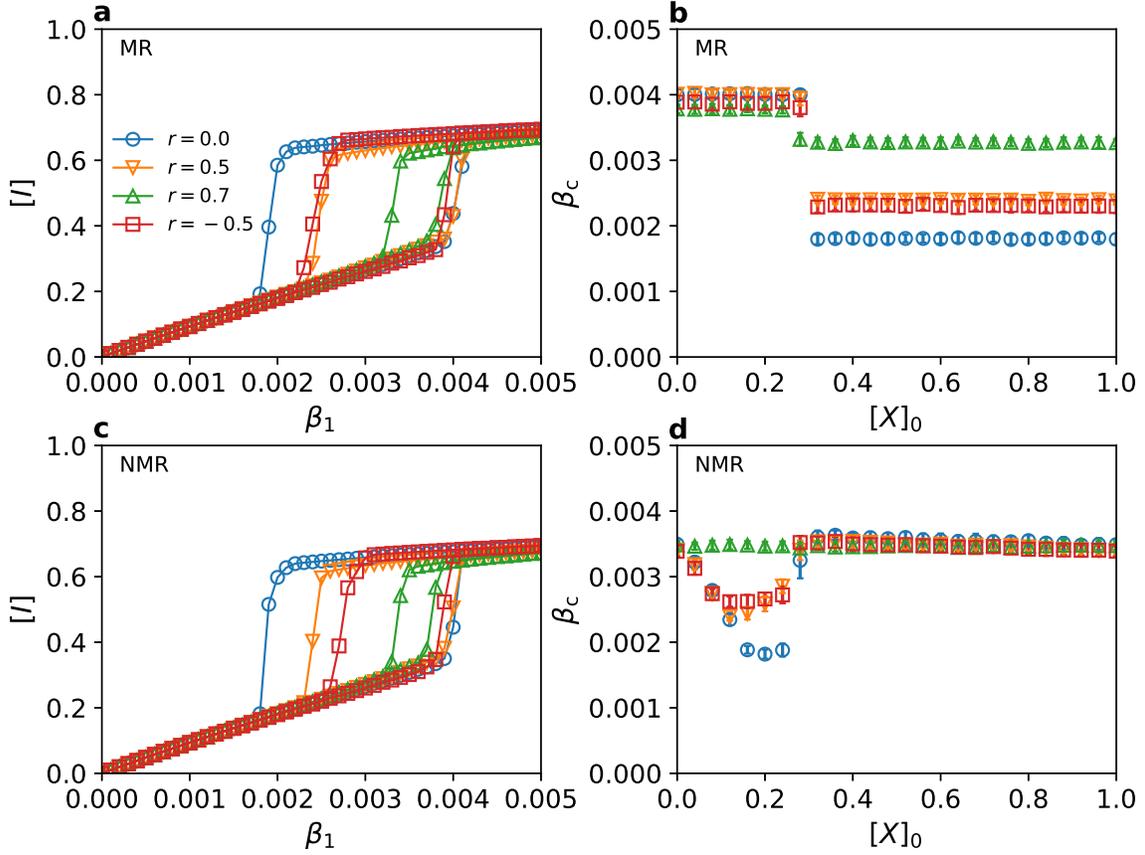}
	\caption{{\bf MR and NMR dynamics in UCNs with degree-degree correlation}.  
(a) Dependence of $[I]$ on $\beta_1$ in the steady
		state for the MR model. The parameter values are $\beta_2=2.0$, $\tau_1=100$
		(corresponding to $\mu_1=0.01$), and $\tau_2=1.0$ (corresponding to
		$\mu_2=1.0$). The threshold of external
		failure of a node is that half of its neighbors have failed. Blue circles,
		orange down triangles, green up triangles and red squares are the simulation
		results for different values of the degree-degree correlation coefficient:
		$r=0$, $r=0.5$, $r=0.7$ and $r=-0.5$, respectively. (b) Dependence of
		$\beta_\textrm{c}$ on the initial value of $[X]_{0}$ for the MR model for $[Y]_{0}=0$.
		(c,d) Simulation results averaged over 50 realizations for the NMR model. The
		networks are of the UCM type with $N=10000$, $k_\textrm{min}=5$, $k_\textrm{max}=100$, and
		different levels of degree-degree correlation.}
	\label{fig:F3}
\end{figure}

\begin{figure} [H]
	\includegraphics[width=\linewidth]{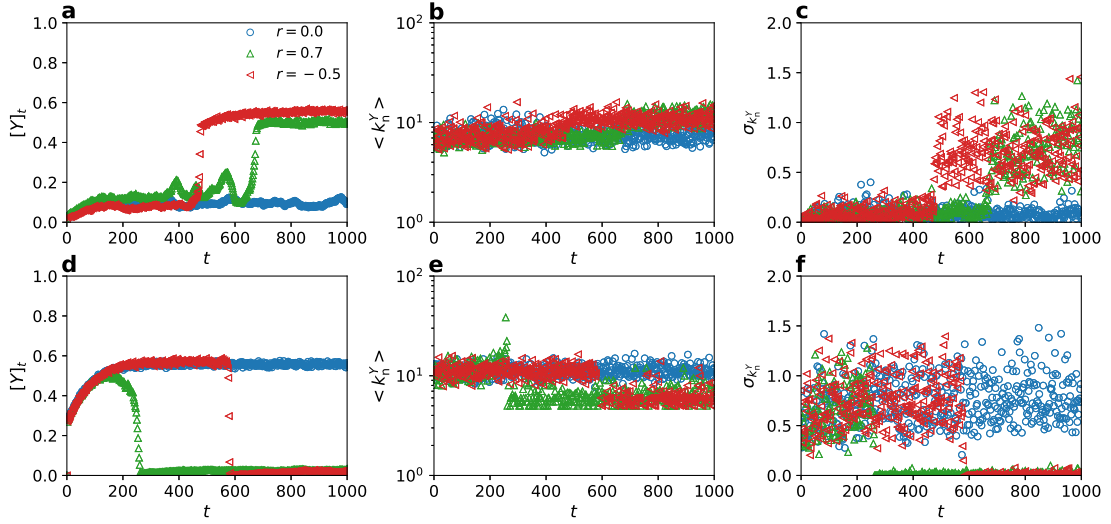}
	\caption{{\bf Dynamical behaviors on UCNs with different levels of degree-degree
		correlation for the MR model}. (a,d) Time evolution of the fraction of $Y$-type
		nodes. Blue circles, green up triangles and red left triangles are for
		$r=0$, $r=0.7$ and $r=-0.5$, respectively. (b,e) Time evolution of the average
		degree of newly emerged $Y$-type nodes. (c,f) Variance of the newly emerged
		Y-type nodes whose degrees correspond to those in Supplementary
		Fig.~\ref{fig:analysis}(b). (a-c): Results for $\beta_1=0.004$ and
		$[X]_0=0.2$. (d-f) Results for $\beta_1=0.0022$ and $[X]_0=0.6$. A single
		realization is used here for better visualization. Other parameter
		values are the same as those in Supplementary Fig.~\ref{fig:F3}.}
	\label{fig:analysis}
\end{figure}

\begin{figure} [H]
	\includegraphics[width=\linewidth]{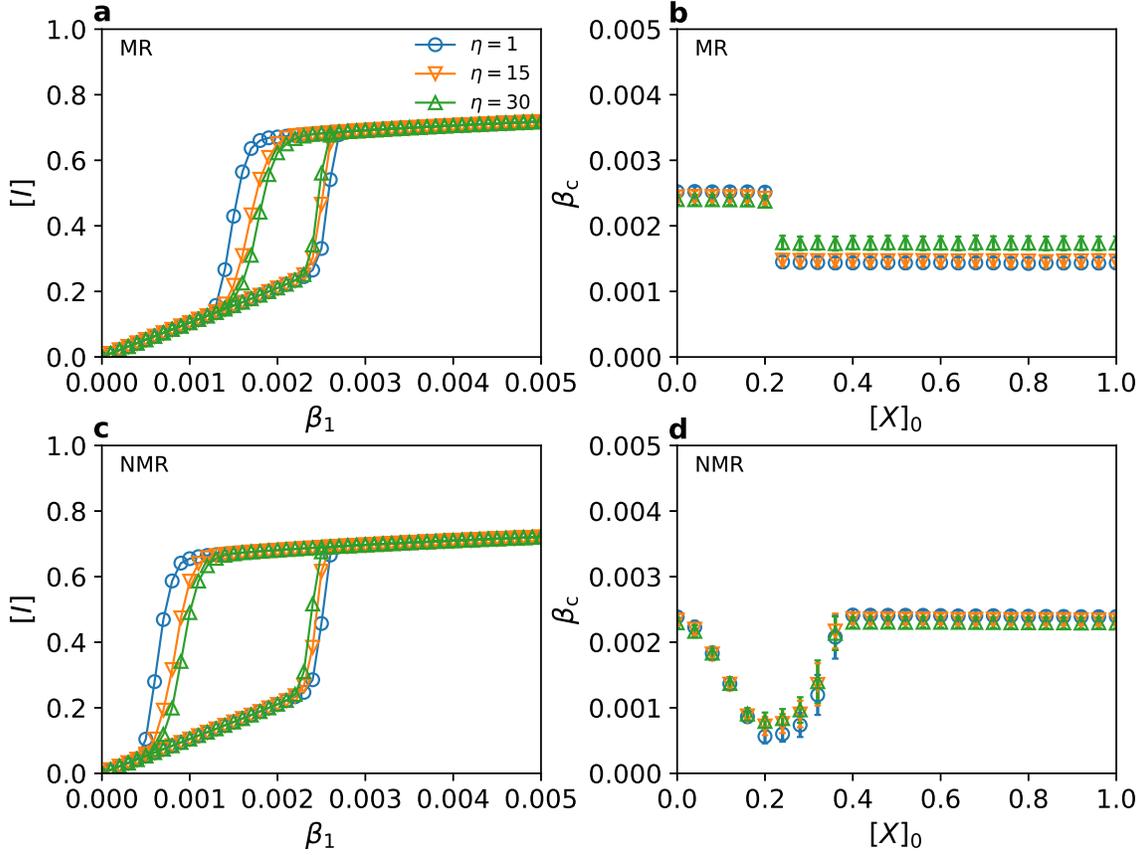}
	\caption{{\bf MR and NMR dynamics in networks with a community structure}. 
(a) Dependence of $[I]$ on $\beta_1$ in the
		steady state for the MR model for $\beta_2=2.4$. Blue circles, orange down
		triangles and green up triangles are simulation results for $Q\approx -0.01$,
		$Q=0.43$ and $Q=0.46$, respectively. Each data point is the result of
		averaging over 20 network realizations, each with 50 random initial
		conditions. (b) Dependence of $\beta_\textrm{c}$ on the initial value of $[X]_{0}$
		for the MR model for $[Y]_{0}=0$. (c,d) The corresponding simulation results
		for the NMR model. The network size is $N=3000$ and the mean degree is
		$\langle k\rangle=6$. Other parameters are the same as those in
		Supplementary Fig.~\ref{fig:F3}.}
	\label{fig:F4}
\end{figure}

\begin{figure} [H]
	\includegraphics[width=\linewidth]{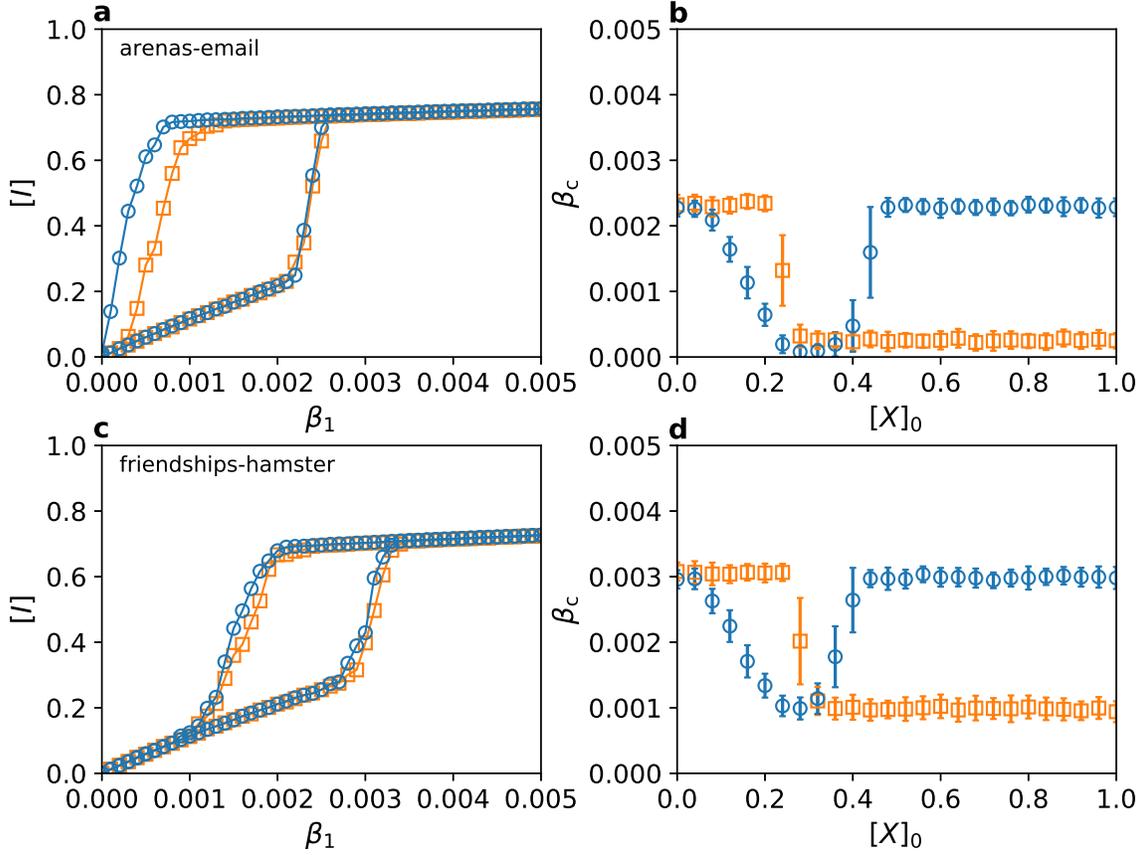}
	\caption{{\bf Comparison between MR and NMR dynamics in the empirical networks}. 
(a) Dependence of $[I]$ on $\beta_1$ in the
		steady state in the arenas-email network for $\beta_2=2.9$. Blue circles
		and orange squares are simulation results averaged over 50 realizations for
		the MR and NMR dynamical models, respectively. (b) Dependence of $\beta_\textrm{c}$
		on the initial value of $[X]_{0}$ for the MR and NMR models for $[Y]_{0}=0$.
		(c,d) The corresponding results in the friendship-hamster network for
		$\beta_2=2.5$. Other parameters are the same as those in Supplementary
		Fig.~\ref{fig:F3}.}
	\label{fig:F5}
\end{figure}

\begin{figure} [H]
	\includegraphics[width=\linewidth]{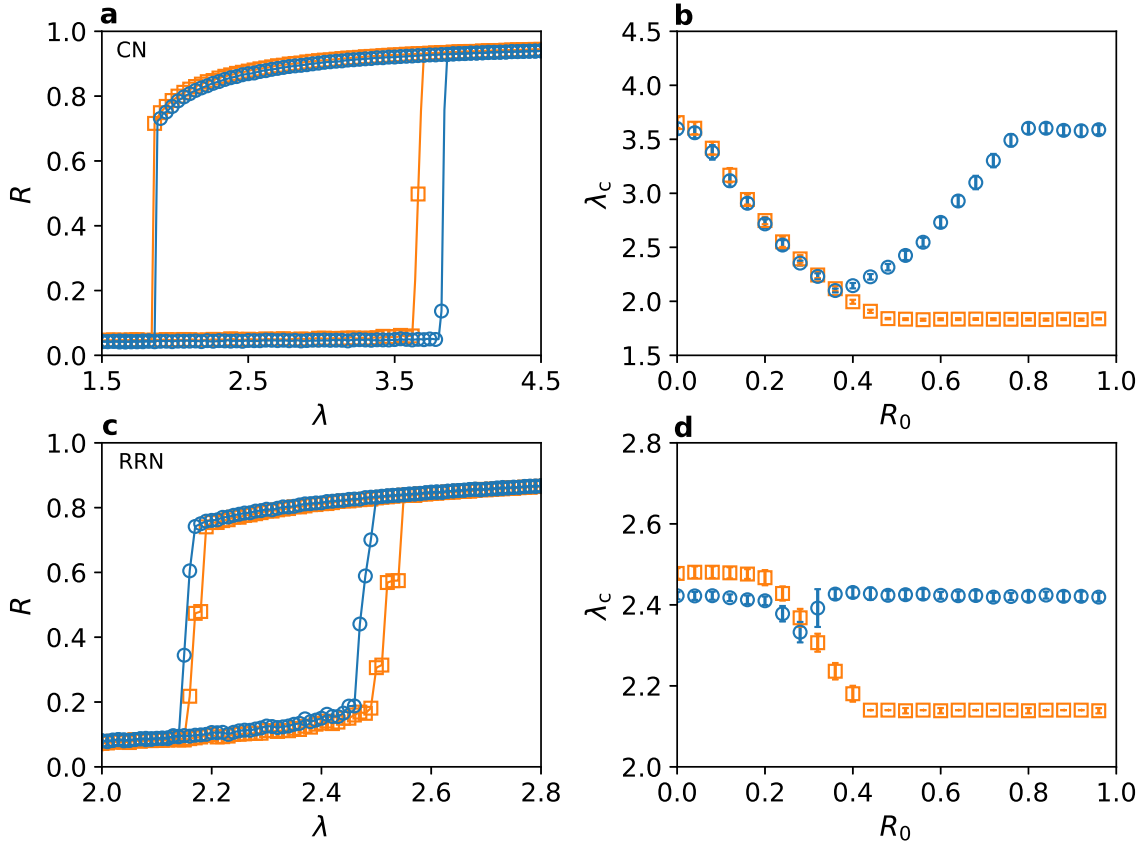}
	\caption{{\bf MR and NMR dynamics for power-grid synchronization}. 
(a,c) Dependence of order parameter $R$ on the coupling strength $\lambda$
		in the steady state for a completely connected network (CN) and a random
		regular network (RRN), respectively. The results are averaged over 20
		realizations. (b,d) Dependence of $\lambda_\textrm{c}$ on the initial order parameter
		$R_0$ for the CN and RRN, respectively. Blue circles and orange squares are
		for the NMR and MR model, respectively. Other parameters are $a=0.05$,
		$p=0.01$, $\omega_0=0$, $\Delta=0.5$, $\phi=0$, $\psi=\pi/4$, $N=5000$, and
		$\langle k\rangle=10$ (for RRN).}
	\label{fig:syn}
\end{figure}

\section{Supplementary Notes}
\subsection{Supplementary Note 1: Pairwise approximation theory}

\subsubsection{\rm Markovian recovery model}

We use symbols of the forms $[U]_t$ and $[UV]_t$ with $U, V \in \{A, X, Y\}$
to denote the fractions of nodes and edges in different states at time $t$,
respectively. For example, $[A]_t$, $[X]_t$ and $[Y]_t$ represent the
fractions of active nodes, $X$-type and $Y$-type failed nodes
at time $t$, respectively, whereas $[AX]_t$ stands for the fraction of active
nodes connected with an $X$-type failed node (i.e., the fraction of $AX$-type
links) at time $t$. Taking into account every link from every node, the
link fractions satisfy the ``conservation law'':
$\sum_{U,V \in \{A,X,Y\}} [UV]_{t} =1$,
with $[UV]_t=[VU]_t$ for $U \neq V$. The supplementary evolution equations for the
fractions of various types of nodes and edges are given by
\begin{equation} \label{eq:mfr_a}
\frac{d[A]_{t}}{dt}=\mu_{1}[X]_{t}+\mu_{2}[Y]_{t}-(\beta_{1}+\beta_{2}E_t)[A]_{t},
\end{equation}
\begin{equation}
\frac{d[X]_{t}}{dt}=\beta_{1}[A]_{t}-\mu_{1}[X]_{t},
\end{equation}
\begin{equation}
\frac{d[Y]_{t}}{dt}=\beta_{2}E_t[A]_{t}-\mu_{2}[Y]_{t},
\end{equation}
\begin{widetext}
	\begin{equation}
	\frac{d[AX]_{t}}{dt}=\mu_{1}[XX]_{t}+\mu_{2}[YX]_{t}+\beta_{1}[AA]_{t}-\mu_{1}[AX]_{t}-(\beta_{1}+\beta_{2}E^{'}_{x,t})[AX]_{t},
	\end{equation}
	\begin{equation}
	\frac{d[AY]_{t}}{dt}=\mu_{1}[XY]_{t}+\mu_{2}[YY]_{t}+\beta_{2}E^{''}_{t}[AA]_{t}-\mu_{2}[AY]_{t}-(\beta_{1}+\beta_{2}E^{'}_{y,t})[AY]_{t},
	\label{eq:mfr_ay}
	\end{equation}
	\begin{equation}
	\frac{d[AA]_{t}}{dt}=\mu_{1}([XA]_{t}+[AX]_{t})+\mu_{2}([YA]_{t}+[AY]_{t})-2(\beta_{1}+\beta_{2}E^{''}_{t})[AA]_{t},
	\end{equation}
\end{widetext}
\begin{equation} \label{eq:mfr_yy}
\frac{d[YY]_{t}}{dt}=\beta_{2}E^{'}_{y,t}([AY]_{t}+[YA]_{t})-2\mu_{2}[YY]_{t},
\end{equation}
\begin{equation} \label{eq:mfr_xx}
\frac{d[XX]_{t}}{dt}=\beta_{1}([AX]_{t}+[XA]_{t})-2\mu_{1}[XX]_{t},
\end{equation}
\begin{equation} \label{eq:mfr_xy}
\frac{d[XY]_{t}}{dt}=\beta_{1}[AY]_{t}+\beta_{2}E^{'}_{x,t}[XA]_{t}-(\mu_{1}+\mu_{2})[XY]_{t},
\end{equation}
where
\begin{equation} \label{eq:mfr_b_e}
E_t=\sum_{j=0}^{m}{C_{k}^{k-j}}\left(\frac{[AI]_{t}}{[A]_{t}}\right)^{k-j}\left(1-\frac{[AI]_{t}}{[A]_{t}}\right)^{j},
\end{equation}
\begin{equation} \label{eq:mfr_b_ex}
{E^{'}_{x,t}}=\sum_{j=0}^{m}{C_{k-1}^{k-1-j}}\left(\frac{[IAX]_{t}}{[AX]_{t}}\right)^{k-1-j}\left(1-\frac{[IAX]_{t}}{[AX]_{t}}\right)^{j},
\end{equation}
\begin{equation} \label{eq:mfr_b_ey}
{E^{'}_{y,t}}=\sum_{j=0}^{m}{C_{k-1}^{k-1-j}}\left(\frac{[IAY]_{t}}{[AY]_{t}}\right)^{k-1-j}\left(1-\frac{[IAY]_{t}}{[AY]_{t}}\right)^{j},
\end{equation}
\begin{equation} \label{eq:mfr_b_eee}
{E^{''}_{t}}=\sum_{j=0}^{m-1}{C_{k-1}^{k-1-j}}\left(\frac{[IAA]_{t}}{[AA]_{t}}\right)^{k-1-j}\left(1-\frac{[IAA]_{t}}{[AA]_{t}}\right)^{j}.
\end{equation}
In the supplementary evolution equation of $[X]_t$ ($[Y]_t$), the first term represents
the fraction of failed $A$-type nodes due to an internal (external) mechanism,
which increases the fraction of the $X$-type ($Y$-type) nodes. The second term
describes the transition that $X$-type ($Y$-type) nodes recover, which
decreases the fraction of the $X$-type ($Y$-type) nodes.

In the supplementary equation of $[AX]_t$, the first (second) term represents the transition
that an $X$-type ($Y$-type) node connected with an $X$-type failed node
recovers spontaneously (i.e., become again an $A$-type node), which increases
the fraction of $AX$ edges. The third term represents the situation that
the A-type neighbors of the $A$-type nodes fail due to an internal mechanism,
which increases the fraction of $AX$ edges. The forth term describes the
transition that $X$-type nodes at the ends of $AX$ edges recover
spontaneously, which decreases the fraction of $AX$ edges. The last term
represents the transition that $A$-type nodes at the ends of $AX$ edges
fail due to an internal or an external mechanism, which decreases the
fraction of $AX$ edges. In the supplementary equation of $[AA]_t$, the first (second) term
represents the transition that $X$-type ($Y$-type) nodes at the ends of
$AX$ and $XA$ ($AY$ and $YA$) edges recover spontaneously, which increases the
fraction of $AA$ edges. The third term represents the transition that
$A$-type nodes at the ends of $AA$ edges fail due to the an internal or an
external cause (i.e., become $X$-type or $Y$-type nodes), which decreases
the fraction of $AA$ edges.

From the set of supplementary equations, we see that, under different conditions, the
probabilities of an active node satisfying the threshold condition are
different. We use the notations $E_t$, $E^{'}_{x,t}$, $E^{'}_{y,t}$ and
$E_t^{''}$ for various probabilities: $E_t$ is the probability for an
active node to satisfy the threshold condition, $E^{'}_{x,t}$ ($E^{'}_{y,t}$)
is the probability for an $A$-type node associated with an $AX$ ($AY$)
edge to satisfy the threshold condition, and $E_t^{''}$ denotes the
probability that an $A$-type node connected with an $A$-type node satisfies
the threshold condition. For all the probabilities, the threshold condition
is $n_A\leq{m}$, where $n_A$ is the number of active neighbors.

Using the pairwise approximation
\begin{equation}
[UVW]_t=\frac{[UV]_t[VW]_t}{[V]_t},
\end{equation}
we have
\begin{equation}
\frac{[IAX]_t}{[AX]_t}=\frac{[AI]_t}{[A]_t},
\end{equation}
\begin{equation}
\frac{[IAY]_t}{[AY]_t}=\frac{[AI]_t}{[A]_t},
\end{equation}
\mbox{and}
\begin{equation}
\frac{[IAA]_t}{[AA]_t}=\frac{[AI]_t}{[A]_t},
\end{equation}
i.e., $E^{'}_{x,t}=E^{'}_{y,t}$. Letting $E^{'}_{x,t}=E^{'}_{y,t}=E^{'}_{t}$,
we have
\begin{equation} \label{eq:mfr_a_e}
E_{t}=\sum_{j=0}^{m}{C_{k}^{k-j}}\left(\frac{[AI]_{t}}{[A]_{t}}\right)^{k-j}\left(1-\frac{[AI]_{t}}{[A]_{t}}\right)^{j},
\end{equation}
\begin{equation} \label{eq:mfr_a_ee}
{E^{'}_{t}}=\sum_{j=0}^{m}{C_{k-1}^{k-1-j}}\left(\frac{[AI]_{t}}{[A]_{t}}\right)^{k-1-j}\left(1-\frac{[AI]_{t}}{[A]_{t}}\right)^{j},
\end{equation}
and
\begin{equation} \label{eq:mfr_a_eee}
{E^{''}_{t}}=\sum_{j=0}^{m-1}{C_{k-1}^{k-1-j}}\left(\frac{[AI]_{t}}{[A]_{t}}\right)^{k-1-j}\left(1-\frac{[AI]_{t}}{[A]_{t}}\right)^{j}.
\end{equation}
Altogether, in the MR model, Supplementary
Eqs.~(\ref{eq:mfr_a}-\ref{eq:mfr_xy}) describe the failure propagation
dynamics.

\subsubsection{\rm Non-Markovian recovery model}

To capture the memory effect of the NMR process, we write the model in terms
of supplementary difference equations by decomposing the NMR process into a series of
MR processes. In the following, each supplementary difference equations describes the
relationship among the fractions of nodes or edges in different states at
time $t+\Delta{t}$ and time $t$. We invoke the notations $[U]_t$ and $[UV]_t$
with $U, V\in\{A, X, Y\}$ to represent the fractions of nodes and
edges of different types at time $t$, respectively. In addition, we use the
notations $[U^l]_t$, $[U^{l_1}V^{l_2}]_t$ and $[U^{l}V]_t$, where $l$, $l_1$
and $l_2$ represent the passing time of the corresponding nodes being in the
current state at time $t$. Due to symmetry, we have $[AX]_t=[XA]_t$. The
supplementary evolutionary equations of the NMR model are given by
\begin{equation} \label{eq:nmfr_b_a}
[A]_{t+\Delta{t}}=[X^{\tau_{1}}]_{t}+[Y^{\tau_{2}}]_{t}+(1-\beta_{1}\Delta{t}-\beta_{2}\Delta{t}E_t)[A]_{t},
\end{equation}
\begin{equation} \label{eq:nmfr_b_x}
[X^{l}]_{t+\Delta{t}}=\left\{
\begin{aligned}
&\beta_{1}\Delta{t}[A]_{t}, & & {{l}\in[0,\Delta{t});}\\
&[X^{l-\Delta{t}}]_{t}, & & {{l}\in[\Delta{t},\tau_{1}]};\\
&0, & & {l\in(\tau_{1},\infty)},
\end{aligned}
\right.
\end{equation}
\begin{equation} \label{eq:nmfr_b_y}
[Y^{l}]_{t+\Delta{t}}=\left\{
\begin{aligned}
&\beta_{2}\Delta{t}E_t[A]_{t}, & & {{l}\in[0,\Delta{t});}\\
&[Y^{l-\Delta{t}}]_{t}, & & {{l}\in[\Delta{t},\tau_{2}]};\\
&0, & & {l\in(\tau_{2},\infty)},
\end{aligned}
\right.
\end{equation}
\begin{widetext}
	\begin{equation} \label{eq:nmfr_b_ax}
	[AX^{l}]_{t+\Delta{t}}=\left\{
	\begin{aligned}
	&\beta_{1}\Delta{t}[AA]_{t}+\beta_{1}\Delta{t}([X^{\tau_{1}}A]_{t}+[Y^{\tau_{2}}A]_{t}), & & {{l}\in[0,\Delta{t});}\\
	&[X^{\tau_1}X^{l-\Delta{t}}]_{t}+[Y^{\tau_2}X^{l-\Delta{t}}]_{t}+(1-\beta_{1}\Delta{t}-\beta_{2}\Delta{t}{E^{'}_{x,t}})[AX^{l-\Delta{t}}]_{t}, & & {{l}\in[\Delta{t},\tau_{1}]};\\
	&0, & & {l\in(\tau_{1},\infty)},
	\end{aligned}
	\right.
	\end{equation}
	\begin{equation} \label{eq:nmfr_b_ay}
	[AY^{l}]_{t+\Delta{t}}=\left\{
	\begin{aligned}
	&\beta_{2}\Delta{t}{E^{''}_{t}}[AA]_{t}+\beta_{2}\Delta{t}{E^{'}_{y,t}}[Y^{\tau_2}A]_{t}+\beta_{2}\Delta{t}{E^{'}_{x,t}}[X^{\tau_1}A]_{t}, & & {{l}\in[0,\Delta{t});}\\
	&[X^{\tau_1}Y^{l-\Delta{t}}]_{t}+[Y^{\tau_2}Y^{l-\Delta{t}}]_{t}+(1-\beta_{1}\Delta{t}-\beta_{2}\Delta{t}{E^{'}_{y,t}})[AY^{l-\Delta{t}}]_{t}, & & {{l}\in[\Delta{t},\tau_{2}]};\\
	&0, & & {l\in(\tau_{2},\infty)},
	\end{aligned}
	\right.
	\end{equation}
\begin{equation} \label{eq:nmfr_b_aa}
\begin{split}
	[AA]_{t+\Delta{t}}=&(1-\beta_{1}\Delta{t}-\beta_{2}\Delta{t}{E^{'}_{t}})([X^{\tau_{1}}A]_{t}
	+[AX^{\tau_{1}}]_{t}+[Y^{\tau_{2}}A]_{t}+[AY^{\tau_{2}}]_{t})\\
	&+\left([X^{\tau_1}X^{\tau_1}]_{t}+[Y^{\tau_2}Y^{\tau_2}]_{t}
	+[X^{\tau_1}Y^{\tau_2}]_{t}+[Y^{\tau_2}X^{\tau_1}]_{t}\right)\\
	&+(1-2\beta_{1}\Delta{t}-2\beta_{2}\Delta{t}{E^{''}_{t}})[AA]_{t},
	\end{split}
	\end{equation}
	\begin{equation} \label{eq:nmfr_b_yy}
	[Y^{l_{1}}Y^{l_{2}}]_{t+\Delta{t}}=\left\{
	\begin{aligned}
	&0, & & {l_{1}\in[0,\Delta{t})\;and\;l_{2}\in[0,\Delta{t})};\\
	&\beta_{2}\Delta{t}{E^{'}_{y,t}}[AY^{l_{2}-\Delta{t}}]_{t}, & & {l_{1}\in[0,\Delta{t})\;and\;l_{2}\in[\Delta{t},\tau_{2}]};\\
	&\beta_{2}\Delta{t}{E^{'}_{y,t}}[Y^{l_{1}-\Delta{t}}A]_{t}, & & {l_{2}\in[0,\Delta{t})\;and\;l_{1}\in[\Delta{t},\tau_{2}]};\\
	&[Y^{l_{1}-\Delta{t}}Y^{l_{2}-\Delta{t}}]_{t}, & & {l_{1}\;and\;l_{2}\in[\Delta{t},\tau_{2}]};\\
	&0, & & {l_{1}\;or\;l_{2}\in(\tau_{2},\infty)},
	\end{aligned}
	\right.
	\end{equation}
	\begin{equation} \label{eq:nmfr_b_xx}
	[X^{l_{1}}X^{l_{2}}]_{t+\Delta{t}}=\left\{
	\begin{aligned}
	&0, & & {l_{1}\in[0,\Delta{t})\;and\;l_{2}\in[0,\Delta{t})};\\
	&\beta_{1}\Delta{t}[AX^{l_{2}-\Delta{t}}]_{t}, & & {l_{1}\in[0,\Delta{t})\;and\;l_{2}\in[\Delta{t},\tau_{1}]};\\
	&\beta_{1}\Delta{t}[X^{l_{1}-\Delta{t}}A]_{t}, & & {l_{2}\in[0,\Delta{t})\;and\;l_{1}\in[\Delta{t},\tau_{1}]};\\
	&[X^{l_{1}-\Delta{t}}X^{l_{2}-\Delta{t}}]_{t}, & & {l_{1}\;and\;l_{2}\in[\Delta{t},\tau_{1}]};\\
	&0, & & {l_{1}\;or\;l_{2}\in(\tau_{1},\infty)},
	\end{aligned}
	\right.
	\end{equation}
	\begin{equation} \label{eq:nmfr_b_xy}
	[X^{l_{1}}Y^{l_{2}}]_{t+\Delta{t}}=\left\{
	\begin{aligned}
	&0, & & {{l_{1}\in[0,\Delta{t})}\;and\;{l_{2}\in[0,\Delta{t})}};\\
	&\beta_{1}\Delta{t}[AY^{l_{2}-\Delta{t}}]_{t}, & & {l_{1}\in[0,\Delta{t})\;and\;l_{2}\in[\Delta{t},\tau_{2}]};\\
	&\beta_{2}\Delta{t}{E^{'}_{x,t}}[X^{l_{1}-\Delta{t}}A]_{t}, & & {l_{2}\in[0,\Delta{t})\;and\;l_{1}\in[\Delta{t},\tau_{1}]};\\
	&[X^{l_{1}-\Delta{t}}Y^{l_{2}-\Delta{t}}]_{t}, & & {l_{1}\in[\Delta{t},\tau_{1}]\;and\;l_{2}\in[\Delta{t},\tau_{2}]};\\
	&0, & & {{l_{1}\in(\tau_{1},\infty)}\;or\;{l_{2}\in(\tau_{2},\infty)}},
	\end{aligned}
	\right.
	\end{equation}
\end{widetext}
where
\begin{equation} \label{eq:nmfr_b_e}
E_{t}=\sum_{j=0}^{m}{C_{k}^{k-j}}\left(\frac{[AI]_{t}}{[A]_{t}}\right)^{k-j}\left(1-\frac{[AI]_{t}}{[A]_{t}}\right)^{j},
\end{equation}
\begin{equation} \label{eq:nmfr_b_eex}
{E^{'}_{x,t}}=\sum_{j=0}^{m}{C_{k-1}^{k-1-j}}\left(\frac{[IAX]_{t}}{[AX]_{t}}\right)^{k-1-j}\left(1-\frac{[IAX]_{t}}{[AX]_{t}}\right)^{j},
\end{equation}
\begin{equation} \label{eq:nmfr_b_eey}
{E^{'}_{y,t}}=\sum_{j=0}^{m}{C_{k-1}^{k-1-j}}\left(\frac{[IAY]_{t}}{[AY]_{t}}\right)^{k-1-j}\left(1-\frac{[IAY]_{t}}{[AY]_{t}}\right)^{j},
\end{equation}
and
\begin{equation} \label{eq:nmfr_b_eee}
{E^{''}_{t}}=\sum_{j=0}^{m-1}{C_{k-1}^{k-1-j}}\left(\frac{[IAA]_{t}}{[AA]_{t}}\right)^{k-1-j}\left(1-\frac{[IAA]_{t}}{[AA]_{t}}\right)^{j}.
\end{equation}
In addition, we have
\begin{equation}
[AX^{l}]_{t}=[X^{l}A]_{t},
\end{equation}
\begin{equation}
[AY^{l}]_{t}=[Y^{l}A]_{t},
\end{equation}
\begin{equation}
[X^{l_{1}}Y^{l_{2}}]_{t}=[Y^{l_{2}}X^{l_{1}}]_{t},
\end{equation}
\begin{equation}
[AI]_{t}=\sum_{l=0}^{\tau_{1}}[AX^{l}]_{t}+\sum_{l=0}^{\tau_{2}}[AY^{l}]_{t},
\end{equation}
\begin{equation}
[AI^{l}]_{t}=[AX^{l}]_{t}+[AY^{l}]_{t}.
\end{equation}
In the supplementary evolutionary equations of $[X^l]_{t+\Delta{t}}$, the fraction of
$X$-type nodes with $l\in[0,\Delta{t})$ at time $t+\Delta{t}$ is equal to
the fraction of $A$-type nodes that fail due to internal causes in the time
interval $[t,t+\Delta{t})$. The fraction of $X$-type nodes with
$\Delta{t}\leq{l}\leq\tau_1$ is equal to the fraction of $X$-type nodes
with $\Delta{t}\leq{l-\Delta{t}}\leq\tau_1$ at time $t$. Since $l$ cannot
exceed the recovery time, the fraction of $X$-type nodes with $l>\tau_1$ is
zero. The supplementary equations for $[Y^l]_{t+\Delta{t}}$ can be obtained in a similar way.

For the supplementary equations of $[AX^l]_{t +\Delta{t}}$, for $l\in[0,\Delta{t})$, the
first term means that the $A$-type neighbors of $A$-type nodes fail due
to internal causes, which increases the fraction of $AX^l$ edges with
$l\in[0,\Delta{t})$ at time $t +\Delta{t}$. The second (third) term depicts
that the $A$-type nodes connected with an $X$-type ($Y$-type) neighbor fail due
to internal causes while their $X$-type ($Y$-type) neighbors recover
spontaneously (i.e., recover because the recovery time has been reached),
which increases the fraction of $AX^l$ edges with $l\in[0,\Delta{t})$. For
$\Delta{t}\leq{l}\leq\tau_1$, the first (second) term describes that the
$X$-type nodes connected with an $X$-type ($Y$-type) neighbor recover
spontaneously, which increases the fraction of $AX^l$ edges, with
${{l}\in[\Delta{t},\tau_{1}]}$ at time $t +\Delta{t}$. The third term
stipulates that there must be no change in the states of $AX^{l-\Delta{t}}$
edges, i.e., the $A$-type nodes associated with $AX^{l-\Delta{t}}$ edges
have not failed during the time interval $[t,t+\Delta{t})$. For $l>\tau_1$,
we have $[AX^l]_{t+\Delta{t}} = 0$.

In the supplementary equation of $[AA]_{t+\Delta{t}}$, the first term denotes that,
associated with edges $X^{\tau_1}A$, $AX^{\tau_1}$, $Y^{\tau_2}A$ and
$AY^{\tau_2}$, the states of $A$-type nodes are not changed, but the states
of the failed nodes have changed, which increases the fraction of $AA$ edges
at time $t +\Delta{t}$. The second term describes that both nodes at the end
of $X^{\tau_1}X^{\tau_1}$, $Y^{\tau_2}Y^{\tau_2}$, $X^{\tau_1}Y^{\tau_2}$,
and $Y^{\tau_2}X^{\tau_1}$ edges recover, which increases the fraction of
$AA$ edges. The third term stipulates that the states of both nodes at the
end of $AA$ edges must not change.

Similar to the MR model, we use $E_t$, $E_t^{'}$, and $E_t^{''}$ to represent
the probabilities that an active node satisfies the threshold condition
$n\leq{m}$ in different cases.

Using the PA $[UVW]_t=[UV]_t[VW]_t/[V]_t$, we have
\begin{equation}
\frac{[IAX]_t}{[AX]_t}=\frac{[AI]_t}{[A]_t},
\end{equation}
\begin{equation}
\frac{[IAY]_t}{[AY]_t}=\frac{[AI]_t}{[A]_t}
\end{equation}
\mbox{and}
\begin{equation}
\frac{[IAA]_t}{[AA]_t}=\frac{[AI]_t}{[A]_t},
\end{equation}
i.e., $E^{'}_{x,t}=E^{'}_{y,t}$. Letting $E^{'}_{x,t}=E^{'}_{y,t}=E^{'}_{t}$,
we have
\begin{equation}
E_{t}=\sum_{j=0}^{m}{C_{k}^{k-j}}\left(\frac{[AI]_{t}}{[A]_{t}}\right)^{k-j}\left(1-\frac{[AI]_{t}}{[A]_{t}}\right)^{j},
\end{equation}
\begin{equation}
{E^{'}_{t}}=\sum_{j=0}^{m}{C_{k-1}^{k-1-j}}\left(\frac{[AI]_{t}}{[A]_{t}}\right)^{k-1-j}\left(1-\frac{[AI]_{t}}{[A]_{t}}\right)^{j},
\end{equation}
and
\begin{equation}
{E^{''}_{t}}=\sum_{j=0}^{m-1}{C_{k-1}^{k-1-j}}\left(\frac{[AI]_{t}}{[A]_{t}}\right)^{k-1-j}\left(1-\frac{[AI]_{t}}{[A]_{t}}\right)^{j}.
\end{equation}

The number of supplementary equations depends on the time step $\Delta t$. If $\Delta t$
is small compared with other time scales in the dynamics, the number of supplementary equations will be large.

\subsubsection{\rm Relationship between MR and NMR models}

For the MR model, the supplementary mean-field equations can be written concisely as
\begin{equation} \label{eq:mfr_mf_x_y}
\left\{
\begin{aligned}
&\frac{d[X]_{t}}{dt}=\beta_{1}[A]_{t}-\mu_{1}[X]_{t},\\
&\frac{d[Y]_{t}}{dt}=\beta_{2}E_t[A]_{t}-\mu_{2}[Y]_{t},
\end{aligned}
\right.
\end{equation}
where
\begin{equation}
E_t=\sum_{j=0}^{m}{C_{k}^{k-j}}{([I]_{t})}^{k-j}(1-[I]_{t})^{j}.
\end{equation}
For the NMR model, in a compact form, the supplementary mean-field equations are
\begin{equation} \label{eq:nmfr_mf_x_y}
\left\{
\begin{aligned}
&[X]_{t+\Delta{t}}=\beta_{1}\Delta{t}[A]_{t}+[X]_{t}-[X^{\tau_1}]_{t},\\
&[Y]_{t+\Delta{t}}=\beta_{2}\Delta{t}E_t[A]_{t}+[Y]_{t}-[Y^{\tau_2}]_{t},
\end{aligned}
\right.
\end{equation}
where
\begin{equation} \label{eq:nmfr_mf_e}
E_t=\sum_{j=0}^{m}{C_{k}^{k-j}}{([I]_{t})}^{k-j}(1-[I]_{t})^{j}.
\end{equation}
When the system has reached a steady state ($t\rightarrow{\infty}$), we have
$[\dot{A}]_t=0$, $[\dot{X}]_t=0$ and $[\dot{Y}]_t=0$.
Supplementary Eq.~\eqref{eq:nmfr_mf_x_y} can be written as
\begin{equation} \label{eq:nmfr_ss_x_y}
\left\{
\begin{aligned}
&\beta_{1}[A]_{t}-[X^{\tau_1}]_{t}=0,\\
&\beta_{2}E_t[A]_{t}-[Y^{\tau_2}]_{t}=0.
\end{aligned}
\right.
\end{equation}
From Supplementary Eq.~\eqref{eq:nmfr_ss_x_y}, we have
\begin{equation}
[X^{\tau_1}]_t=\beta_1[A]_t=C_1
\end{equation}
\mbox{and}
\begin{equation}
[Y^{\tau_2}]_t=\beta_{2}E[A]_t=C_2,
\end{equation}
where $C_1$ and $C_2$ are constants. In addition, we have
\begin{equation}
[X^l]_t=[X^{\tau_1}]_{t+\tau_1-l}=C_1
\end{equation}
\mbox{and}
\begin{equation}
[Y^l]_t=[Y^{\tau_2}]_{t+\tau_2-l}=C_2.
\end{equation}
We then have
\begin{equation}
[X]_t=\Sigma_l^{\tau_1}[X^l]_t=\tau_1[X^{\tau_1}]_t,
\end{equation}
which gives
\begin{equation}
[X^{\tau_1}]_t=\frac{1}{\tau_1}[X]_t.
\end{equation}
Similarly, we can get
\begin{equation}
[Y^{\tau_2}]_t=\frac{1}{\tau_2}[Y]_t.
\end{equation}
Supplementary Equation~\eqref{eq:nmfr_ss_x_y} can then be rewritten as
\begin{equation} \label{eq:nmfr_ss2_x_y}
\left\{
\begin{aligned}
&\beta_{1}[A]_{t}-\frac{1}{\tau_1}[X]_{t}=0,\\
&\beta_{2}E_t[A]_{t}-\frac{1}{\tau_2}[Y]_{t}=0.
\end{aligned}
\right.
\end{equation}
Comparing with the steady-state solution of the MR model:
\begin{equation} \label{eq:mfr_ss2_x_y}
\left\{
\begin{aligned}
&\beta_{1}[A]_{t}-\mu_1[X]_{t}=0,\\
&\beta_{2}E_t[A]_{t}-\mu_2[Y]_{t}=0,
\end{aligned}
\right.
\end{equation}
we see that the steady states for both models are equivalent to each other
for fixed values of $\mu_1=1/\tau_1$ and $\mu_2=1/\tau_2$.

\subsection{Supplementary Note 2: Effects of different initial conditions on failure propagation and evolutionary trajectories of failed nodes}

Supplementary Figs.~\ref{fig:rho0_betaC} and \ref{fig:rho0_rho} illustrate
the effects of initial conditions on failure propagation dynamics.
The different recovery mechanisms make the competition processes between the
$X$-type and $Y$-type nodes distinct for the MR and NMR models. In
general, non-Markovian features render more complicated the failure-recovery
dynamics. Here we investigate how the fraction of initially failed nodes
$[I]_0$ influences the phase transition in the NMR model.

We set the initial conditions as $[X]_{0} \neq 0$ and $[Y]_{0}=0$. When the
fraction of initially failed nodes is sufficiently small, e.g., $[X]_0=0.05$,
as in Supplementary Fig.~\ref{fig:trajectory_and evolution}(a), active nodes
fail most likely because of internal causes. In both MR and NMR models,
the system will evolve into a low-failure state (phase), where the
trajectories evolve from $t_\textrm{O}$ to $t_\textrm{A}$. As $[X]_0$ increases to $0.2$, as
shown in Supplementary Fig.~\ref{fig:trajectory_and evolution}(d), the
evolutionary process for the MR model is similar to the case in
Supplementary Fig.~\ref{fig:trajectory_and evolution}(a): the trajectory
evolves from $t_\textrm{O}$ to $t_\textrm{A}$. This is because the $X$-type nodes not only are
born at the rate $\beta_1$ but also recover at the rate $\mu_1$, driving the
system to a dynamic equilibrium associated with the low-failure phase. As a
result, there are insufficient newly created $X$-type nodes to cause a
large scale external failure. In contrast, in the NMR model, $X$-type nodes
are always created but, in the early stage $[t_\textrm{O},t_\textrm{A}]$, there is no recovery
because of the memory effect. Once the fraction of failed nodes, including the
initial and the newly created $X$-type nodes in $[t_\textrm{O},t_\textrm{A}]$, reaches a
critical value determined by the criterion that many $A$-type nodes with
active neighbors satisfy the threshold condition $n_A\leq{m}$, $[Y]$ increases
rapidly and the trajectory moves from $t_\textrm{A}$ to $t_{\textrm{A}'}$. In the time
interval $[t_{\textrm{A}'},t_{\textrm{B}'}]$, $[Y]$ decreases slowly due to a short
recovery time $\tau_2=1$, while internal failures make $[X]$ increase
slowly because $t_{\textrm{B}'}<100$. In the time interval $t\in[100,100+\Delta{t})$,
the age of the initially failed nodes ($[I]_0=[X]_0=0.2$) reaches the recovery
time and these nodes will turn into $A$ state simultaneously. Consequently,
$[X]$ decreases suddenly to a low value and the trajectory evolves sharply
from $t_{\textrm{B}'}$ to $t_{\textrm{C}'}$. At this time, $[I]$ still remains at a high
value, making more active nodes (including the original and new $X$-type nodes)
satisfy the threshold condition $n_A\leq{m}$ and resulting in a rapid growth
of $Y$-type nodes in the time interval $[t_{\textrm{C}'},t_{\textrm{D}'}]$. At time
$t_{\textrm{D}'}$, $[I]$ increases to a higher value, rendering more competitive
$Y$ than $X$ state as they compete for $A$-type nodes. As a result, $A$-type
nodes fail due mostly to external causes, and $[Y]$ ($[X]$) continues to
increase (decrease) in the time interval $[t_{\textrm{D}'},t_{\textrm{E}'}]$.

For $[X]_0=0.3$, as shown in Supplementary
Fig.~\ref{fig:trajectory_and evolution}(g),
the evolutionary process of the MR model is consistent with that in
Supplementary Fig.~\ref{fig:trajectory_and evolution}(d). For the NMR model,
the evolutionary process is different from that in
Supplementary Fig.~\ref{fig:trajectory_and evolution}(b). In the time interval
$[t_{\textrm{B}'},t_{\textrm{C}'}]$, more initially failed nodes with $[I]_0=[X]_0=0.3$
recover simultaneously because their age (i.e., the current time) has reached
the recovery time $\tau_1=100$. At time $t_{\textrm{C}'}$, a relatively low value
of $[I]$ makes fewer active nodes satisfy the threshold condition
$n_A\leq{m}$. Because of the short recovery time, all the current $Y$-type
nodes switch into the active state in the time interval $[t_{\textrm{C}'},t_{\textrm{D}'}]$.
After that, many active nodes fail internally and become $X$-type nodes,
so $[X]$ increases continuously and the trajectory moves from $t_{\textrm{D}'}$
to $t_\textrm{A}$.

When the initial value $[X]_0$ is sufficiently large, e.g., $0.6$ as in
Supplementary Fig.~\ref{fig:trajectory_and evolution}(j), many $A$-type nodes
will fail due to external causes and $[Y]$ will increase rapidly in both
models, as shown in the trajectory from $t_\textrm{O}$ to $t_\textrm{A}$. After that, in the MR
model, the increment of $[Y]$ enhances the probability for active nodes to
switch into the $Y$ state, while increasingly $X$-type nodes recover and turn
to the $A$ state. The corresponding trajectory moves from $t_\textrm{A}$ to $t_\textrm{B}$. In
the NMR model, the evolutionary process is similar to that in
Supplementary Fig.~\ref{fig:trajectory_and evolution}(g).

\subsection{Supplementary Note 3: Markovian and non-Markovian dynamics when external recovery is slower than internal recovery}

In the main text, the case where internal recovery is slower than external
recovery, i.e., $\tau_1>\tau_2$, is treated. There are situations in the
real world where the opposite, i.e., $\tau_1<\tau_2$, can occur. For example,
in the case of an earthquake, the recovery of a node in an infrastructure
(e.g., restoring a damaged building) may require more time than that of
repair due to internal material failures. To study this case, we fix the
network topology as in Supplementary Fig.~\ref{fig:rho0_betaC}.
Supplementary Fig.~\ref{fig:F1} shows the representative results. Comparing
Supplementary Fig.~\ref{fig:F1}(a) with Fig. 2(b) in main text, we see that
they demonstrate qualitatively similar behaviors, indicating the applicability
of our theoretical framework for the case $\tau_1<\tau_2$.

Supplementary Fig.~\ref{fig:F2} shows the phase diagram on the
initial-condition plane $([X]_{0},[Y]_{0})$. Comparing it with Fig.~7 in
the main text, we see that both exhibit similar phenomena, although the
quantitative details are different. This further validates that our developed
framework is applicable to the case $\tau_1<\tau_2$.

\subsection{Supplementary Note 4: Effects of network structure on Markovian and
non-Markovian recovery dynamics} \label{Appendix:D}

In general, network topology can have a significant effect on the cascading
process of failure propagation. In the main text, we have discussed two kinds
of network topology, i.e., random regular and scale-free networks, and found
that memory in the nodal recovery can counterintuitively make both types of
networks more resilient against large scale failures. Here we study two
additional types of network: those with degree-degree correlation and a
community structure, respectively, and demonstrate the same effect of memory.

\subsubsection{\rm Effects of degree-degree correlation} \label{Appendix:D1}

To generate networks with adjustable degree-degree correlation coefficients,
we use the standard edge-rewiring procedure~\cite{xulvi2004reshuffling,
gao2016effective}. In particular, we first generate an uncorrelated
configuration network (UCN)~\cite{catanzaro2005generation} with the degree
range $[k_\textrm{min},\sqrt{N}]$ and degree distribution $P(k)\sim k^{-\gamma}$.
Keeping the degree of each node unchanged, we adjust the degree-degree
correlation through the following process. Firstly, at each step, we randomly
choose two edges in the network, disconnect them, and switch the two links
among the four chosen nodes. Secondly, to generate an assortative (a
disassortative) network, we add a new edge between the highest degree and
the second highest (lowest) nodes and then connect the remaining pair of
nodes. If either of the new edges already exists, we leave the network
unchanged. Thirdly, we repeat the process until the observed degree-degree
correlation coefficient has reached a target value, which is defined
as~\cite{newman2018networks}:
\begin{equation} \label{eq:r_coef}
r=\frac{\sum_{ij}(A_{ij}-k_{i}k_{j}/2m_\textrm{e})k_{i}k_{j}}{\sum_{ij}(k_{i}\delta_{ij}-k_{i}k_{j}/2m_\textrm{e})k_{i}k_{j}},
\end{equation}
where $m_\textrm{e}$ is the total number of edges in the network, $A_{ij}= 1$ if there
is an edge between the node $i$ and $j$ (otherwise, $A_{ij}= 0$),
$\delta_{ij}$ is the Kronecker delta ($1$ if $i= j$ and $0$ otherwise).
There is no degree-degree correlation for $r=0$, but the network will
have positive (negative) degree-degree correlation for $r>0$ ($r<0$).

Supplementary Figs.~\ref{fig:aij}(a-d) show the degree-degree correlation
properties of the UCNs for $r=0$, $r=0.5$, $r=0.7$ and $r=-0.5$, respectively,
where it can be visually seen that the high-degree nodes in the network
with a positive correlation tend to be connected together, while a high-degree
node in the case of negative correlation tends to be connected to a low-degree
one. Supplementary Fig.~\ref{fig:dist_knn} shows the probability distribution
of one node of degree $k_i$ connecting to another node of degree $k_j$,
where panels (a-d) correspond to the cases of $k_i=5$, $k_i=20$, $k_i=40$ and
$k_i=60$, respectively.

Supplementary Fig.~\ref{fig:F3} shows the results of spontaneous recovery
dynamics on UCNs with different levels of degree-degree correlation for
$N=10000$, $k_\textrm{min}=5$, and $k_\textrm{max}=100$. Supplementary Figs.~\ref{fig:F3}(a)
and \ref{fig:F3}(c) show the dependence of $[I]$ on adiabatic variations in
$\beta_1$ in the steady state for the MR and NMR models, respectively. The
blue circles, orange down triangles, green up triangles and red squares
are the average simulation results for degree-degree correlation coefficient
$r=0$, $r=0.5$, $r=0.7$ and $r=-0.5$, respectively, where the system
is regarded as in the low-failure phase when $\beta_1$ is smaller than the
critical value $\beta_\textrm{c}$ and a large scale failure occurs when $\beta_1$
exceeds $\beta_\textrm{c}$. For the case of zero degree-degree correlation, the
network can remain in the low-failure phase for the largest possible value
of $\beta_\textrm{c}$, indicating the highest possible degree of resilience. However,
in the high-failure phase with adiabatically decreasing value of $\beta_1$,
the network with a high level of degree-degree correlation has a larger value
of $\beta_\textrm{c}$, signifying a stronger ability to recover from damage. As a
result, in spontaneous recovery models, UCNs with degree-degree correlations
can make the hysteresis region smaller, regardless of positive or negative
degree-degree correlation. Supplementary Figs.~\ref{fig:F3}(b) and
\ref{fig:F3}(d) present further evidence to support the finding in the main
text that non-Markovian recovery makes the networks more resilient against
large scale failures, regardless of the detailed network structure.

To explain why positive or negative degree-degree correlation shrinks the
hysteresis region in Fig.~\ref{fig:F3}, we focus on the MR model.
Supplementary Fig.~\ref{fig:analysis}(a) shows the time evolution of the
fraction of Y-type nodes for $\beta_1=0.004$ and $[X]_0=0.2$, where the
the system is in the low-failure phase. The degree correlation tends to
reduce the resilience of system, thereby promoting cascading failures.
Corresponding to the steady-state of $[I]$ in Supplementary
Fig.~\ref{fig:F3}(a), we see that the large scale failure results from the
sharp increase of $[Y]$. Supplementary Figs.~\ref{fig:analysis}(b) and
\ref{fig:analysis}(c) show the evolution of the mean degree of the newly
emerged Y-type nodes and the standard variance of the nodal degree,
respectively. It can be seen that, for the networks with $r=0.7$ ($-0.5$),
the degree values of many newly emerged Y-type nodes are similar for
$t<650$ ($t<470$). However, after that, the degrees of newly emerged node
vary. While the results of both cases look similar, the underlying topological
properties of the network are different. In particular, in scale-free networks
with a highly heterogeneous degree distribution, a large proportion of the
nodes have smaller degrees than the mean degree, but a few hubs have larger
degrees. For an assortative network, nodes of similar degree are more likely
to be connected together. As a result, there are more vulnerable nodes of
low degrees connecting to each other [e.g., the nodes with $k_i=5$ as shown
in Supplementary Fig.~\ref{fig:dist_knn}(a)], which comprise a sizable
vulnerable component, making the network more susceptible to a large scale
failure~\cite{payne2009information}. When the threshold on external failure
(i.e., the critical fraction of the active neighbors) is fixed (e.g., $0.5$),
a low-degree node is vulnerable because it is more easily affected by its
neighboring nodes. For a disassortative network, the high-degree nodes prefer
to connect to the vulnerable nodes of low degrees [e.g., the nodes of
$k_i=20$ shown in Supplementary Fig.~\ref{fig:dist_knn}(b)], making the hubs
more vulnerable against the failures of their neighboring nodes and promoting
cascading failures.

Supplementary Fig.~\ref{fig:analysis}(d) shows the time evolution of $[Y]_t$
for the MR model for $\beta_1=0.0022$ and $[X]_0=0.6$, corresponding to the
case where the degree correlation suppresses the occurrence of cascading
failure when the system is in the high-failure phase. From Supplementary
Figs.~\ref{fig:analysis}(d-f), we see that for the cases of $r=0.7$ and
$r=-0.5$, there are a large number of externally failed nodes of various
degrees in the transient process, after which the system undergoes a sharp
decrease in $[Y]_t$ near $t=250$ and $t=570$, respectively, leaving few
externally failed nodes with a similar degree. The reason that the fraction
of $[Y]_t$ decreases sharply is also the vulnerable components. For a small
value of $\beta_1$, the recovery of a few vulnerable nodes will cause
successive recovery of the other connected vulnerable nodes and eventually
block failure propagation. Note that, for $r = 0$, the high-failure phase
continues, because a more stable high-failure phase depends on the structure
in which nodes of quite distinct degrees are connected together, as verified by
the results in Supplementary Figs.~\ref{fig:aij} and \ref{fig:analysis}(f).

\subsubsection{\rm Effects of community structure}

For simplicity, we consider networks that contain two two
communities~\cite{girvan2002community,condon2001algorithms}, where each
community is composed of $n$ nodes and so the network size is $N=2n$. Nodes
in the same community are connected to each other with the probability
$p_\textrm{in}$, while links across the two communities occur with the probability
$p_\textrm{out}$, so the ratio of the numbers of edges in and between the communities
is $\eta\approx{p_\textrm{in}/p_\textrm{out}}$. The degree distributions of the subnetwork
in each community and of the whole network are Poisson. Note that the value
of $\eta$ determines the strength of the community structure. In particular,
$\eta=1.0$ means that the numbers of edges in and between communities are
the same so, effectively, there is no community structure. Likewise, a value
of $\eta$ much greater than one indicates a stronger community structure with
significantly more edges within the individual communities. Quantitatively,
the community strength can be measured by the modularity $Q$ defined
as~\cite{newman2018networks}
\begin{equation} \label{eq:q_coef}
Q=\frac{1}{2m_\textrm{e}}{\sum_{ij}(A_{ij}-\frac{k_{i}k_{j}}{2m_\textrm{e}})\delta(c_{i},c_{j})},
\end{equation}
where $m_\textrm{e}$ is the total number of edges in the network, $A_{ij}$ are the
elements of the network adjacency matrix, $c_i$ is the community to which
node $i$ belongs, and $\delta_{ij}$ denotes the Kronecker delta. The value
of $Q$ is strictly less than $1.0$, where $Q>0$ ($Q<0$) means that there are
more (less) edges between nodes in the same communities than can be expected
by chance.

In our numerical simulations, we fix $N=3000$ and mean degree
$\langle k\rangle=6$, and generate three typical networks with $\eta=1.0$,
$15.0$ and $30.0$, with the corresponding approximate $Q$ values
$-0.01$, $0.43$ and $0.46$, respectively. Initially, we randomly distribute
seeds in the whole network. Supplementary Fig.~\ref{fig:F4} demonstrates the
dynamical behaviors for both the MR and NMR models. It can be seen that, if
the system is in a low-failure (high-failure) phase, the community structure
tends to reduce (enhance) resilience. The reason is that nodes in the same
community tend to be connected more closely and thus constitute a
high clustering structure. The failure of a node due to an external mechanism
tends to enhance the probability for other nodes in the same community to
fail. Similarly, if a node has recovered, the probability for other nodes in
the same community to recover is increased. As a result, when the system is in
the low failure phase, the community structure makes a large scale failure
more likely. In contrast, if the system is in the high failure phase, the
community structure can facilitate recovery. Qualitatively, these results are
consistent with those in Supplementary Note 4.

\subsection{Supplementary Note 5: Markovian and non-Markovian recovery dynamics in empirical networks}

We study Markovian and non-Markovian recovery dynamics in empirical networks,
using the arenas-email and the friendship-hamster networks as two examples,
where the former is an email communication
network~\cite{konect:2017:arenas-email} of size $N=1133$ and average degree
$\langle k\rangle=9.6$, and the latter is the network of friendship among
the users of the website
hamsterster.com~\cite{konect:2017:petster-friendships-hamster} with $N=1858$
and $\langle k\rangle=13.4$. Simulation results reveal the same phenomenon
as in other cases, i.e., non-Markovian type of recovery with a memory tends
to enhance the network resilience against large scale failures, as shown in
Supplementary Fig.~\ref{fig:F5}.

\subsection{Supplementary Note 6: Markovian and non-Markovian recovery dynamics in power-grid
synchronization} \label{Appendix:E}

An idealized model for a power grid is the network of Kuramoto oscillators,
where the weighted coupling coefficient between two oscillators is related to
their own natural frequencies~\cite{dorfler2012syn,wang2011syn}. We study
the following model~\cite{zhang2013explosive}:
\begin{equation} \label{eq:Kuramoto_old}
\frac{d\theta_{i}}{dt}=\omega_i+\frac{\lambda|\omega_i|}{k_i}
\sum_{j=1}^NA_{ij}\sin(\theta_j-\theta_i), \ \
i=1,\ldots,N \ ,
\end{equation}
where $\omega_i$ is the natural frequency of oscillator $i$ randomly chosen
from the Lorentzian distribution
\begin{equation}
g(\omega)=\frac{1}{\pi}[\frac{\Delta}{(\omega-\omega_0)^2+\Delta^2}],
\end{equation}
$\lambda$ is the overall coupling strength, $k_i=\sum_{j=1}^NA_{ij}$ is the
degree of node $i$, and $A_{ij}$ are the elements of the symmetric adjacency
matrix. In this model, explosive synchronization~\cite{zhang2013explosive,
hu2014exact} can arise. In the regime of complete synchronization, all
oscillators are localized in the same area. To investigate the propagation of
synchronization under the MR and NMR dynamics, we fix the synchronization
area and construct the following dynamical system of phase oscillators:
\begin{equation} \label{eq:Kuramoto}
\frac{d\theta_{i}}{dt}=\omega_i+\frac{\lambda|\omega_i|}{k_i}
\sum_{j=1}^NA_{ij}\sin(\theta_j-\theta_i)-a\sin(\theta_i-\phi), \ \
i=1,\ldots,N \ ,
\end{equation}
where $a$ is a parameter characterizing the degree of synchronization.
The first and third terms on the right-hand side describe the self-dynamics
of node $i$, while the second term represents the interactions between node
$i$ and its interacting partners. To see the meaning of $\phi$, we introduce
the order parameter $R$
\begin{equation} \label{eq:order}
Re^{i\Psi}=\frac{1}{N}\sum_{j=1}^N e^{i\theta_j},
\end{equation}
where $0\leq R \leq 1$ and $\Psi$ denotes the average phase. The
synchronization region can then be defined as $[\phi-\psi, \phi+\psi]$.
For the MR dynamics, each oscillator in the synchronized region has the
probability $p$ to return to the asynchronous state with a random phase
between $[-\pi, \pi]$. For the NMR process, each oscillator can stay in
the synchronized region for time $\tau=1/p$ before leaving the region
to become asynchronous with others.

We carry out synchronization simulations on a completely connected network (CN)
and a random regular network (RRN) with average degree $\langle k\rangle=10$.
We find that, for both networks, the system transitions to the high-failure
(low-failure) phase when increasing (decreasing) $\lambda$ towards a critical
value $\lambda_\textrm{c}$. Supplementary Fig.~\ref{fig:syn} shows the dynamic
behaviors for both the MR (orange squares) and NMR (blue circles) dynamics.
It can be seen that the results are qualitatively similar to
those in either Supplementary Fig.~\ref{fig:F4} or Supplementary
Fig.~\ref{fig:F5}, indicating that non-Markovian recovery makes the network
more resilient against large scale breakdown of synchronization. The results
also indicate that the network topology can have a considerable impact on the
system resilience.



\def\bibsection{\section{Supplementary References}} 